\documentclass[printer]{aa}
\usepackage{setspace}
\usepackage{amssymb}
\usepackage{amsmath}
\usepackage{sidecap}
\usepackage{mathptmx}
\usepackage{graphicx}
\usepackage{hyperref}

\usepackage{natbib}
\usepackage{rotating}
\usepackage{soul}
\usepackage{url}
\usepackage{epsfig}
\usepackage{longtable}
\usepackage{subcaption}
\usepackage{enumitem}

\usepackage{color}
\begin{document}

\title{Characterizing galaxy clusters by their gravitational potential: systematics of cluster potential reconstruction}
\titlerunning{Systematics of the reconstructions of cluster mass and potential}

\author{C. Tchernin\inst{1,2}, E. T. Lau\inst{3} , S. Stapelberg\inst{1,2}, D. Hug\inst{1,2}, M. Bartelmann\inst{1,2} }
\institute{\inst{1} Institut f\"ur Theoretische Astrophysik, Zentrum f\"ur Astronomie, Heidelberg University, Philosophenweg 12, 69120 Heidelberg, Germany,
\\\inst{2} Institut f\"ur Theoretische Physik, Heidelberg University, Philosophenweg 16, 69120 Heidelberg, Germany,
\\\inst{3} Department of Physics, University of Miami, Coral Gables, FL 33124, U.S.A\\ e-mail: Tchernin@uni-heidelberg.de, }

\authorrunning{C. Tchernin et al.}


\abstract
{Biases in mass measurements of galaxy clusters are one of the major limiting systematics in constraining cosmology with clusters. }
{We aim to demonstrate that the systematics associated with cluster gravitational potentials  are smaller than the hydrostatic mass bias and that cluster potentials could therefore be a good alternative to cluster masses in cosmological studies. }
{Using cosmological simulations of galaxy clusters, we compute the biases in the hydrostatic mass (HE mass) and those in the gravitational potential, reconstructed from measurements at X-ray and millimeter wavelengths.  
In particular, we investigate the effects  of the presence of substructures and of non-thermal pressure support on both the HE mass and the reconstructed potential.
}
{
We find that the bias in the reconstructed potential (6\%) is less than that of the HE mass (13\%), and that the scatter in the reconstructed potential decreases by $\sim 35\%$ with respect to that in the HE mass. 
}
{
This study shows that  characterizing galaxy clusters by their gravitational potential is a promising alternative to using cluster masses in cluster cosmology.
}

\maketitle

\section {Introduction}

Galaxy clusters contain rich astrophysical and cosmological information.  For instance, the statistics of the galaxy cluster population as a function of redshift and mass is often used to trace the evolution of the large-scale structures \citep[\textit{cluster counts}, see e.g.,][]{voit05,planck15_cosmocluster}. Large existing and upcoming missions, such as the Dark Energy Survey \citep[DES,][]{des}, Euclid \citep{laureijs11} and LSST \citep{lsst} aim to use galaxy clusters to study the nature of the dark energy that is responsible for accelerating the expansion of the Universe. However, to be competitive these cluster based cosmological probes require well constrained cluster masses. This is unfortunately difficult to achieve because cluster masses are not directly observable. The tension between the results from two different cosmological probes, cluster counts \citep{planck13_cosmocluster,planck15_cosmocluster} and CMB cosmology \citep{planck13_cosmoCMB,planck15_cosmoCMB} illustrates well this problem, as this discrepancy may be due to the cluster mass estimates \citep[e.g., ][]{planck13_cosmocluster}.

We have nowadays access to high quality multi-wavelength data to study in detail the galaxy cluster components. It is therefore crucial to develop new analysis tools to explore the rich content of these data sets. We demonstrate in this paper that cluster potentials can be reconstructed with higher accuracy than the hydrostatic mass (hereafter, HE mass) from cluster observations, regardless of the dynamical state of the cluster.

Clusters are bright diffuse X-ray sources that emit Bremsstrahlung and spectral lines. The observations of this X-ray emission provide information on the electronic pressure, temperature, density and composition of the intracluster medium (ICM)  \citep[see e.g.,][]{eckert18}. Due to the steep density profile in galaxy clusters and the dependence of the Bremsstrahlung radiation on the squared electron density \citep[e.g., ][]{sarazin88}, X-ray observations are mostly sensitive to the central regions of the clusters (within $\text{R}_{500} \sim 0.6 \text{R}_{200}$).
The ICM gas can also be observed at millimeter wavelengths via its thermal Sunyaev-Zel'dovich (tSZ) effect. This signal appears as a distortion of the unperturbed cosmic microwave background (CMB) spectrum and arises when the thermal electrons of the ICM gas interact via inverse Compton scattering with the CMB photons. The magnitude of this distortion is proportional to the thermal electronic pressure integrated along the line of sight \citep[e.g.,][]{SZ69,sz,sayers11} and provides information on the ICM gas at large distance from the cluster center \citep[up to about $\text{R}_{200}$, see e.g.][]{planck13}.
In addition, the line-of-sight projected (2D) gravitational potential can be inferred from gravitational lensing \citep[GL, see][for a review]{bartelmann10}. This is a purely geometric effect that probes the total matter density (the baryonic matter and the DM) projected along the line-of-sight without requiring any knowledge on the dynamical state of the matter. We consider here two regimes: strong and weak lensing. Gravitational arcs and multiple-image systems of a source are strong lensing features that appear in the cluster core, where the density is largest \citep[up to $\sim$ 0.2$\text{R}_{500}$, e.g.,][]{jauzac19,mahler19}.  In contrast, weak lensing features are small deformations of background galaxies  that can only be statistically analyzed \citep[see][for a review]{bartelmann01}  and probes regions located further away from the cluster center \citep[up to about 2$\text{R}_{200}$, e.g.,][]{linden14b,klein19,umetsu20}.
The combination of these multi-wavelength observations allows us to study the galaxy cluster thermodynamic properties over a wide spatial range \citep[e.g.,][]{tchernin16,sh18,eckert18,strait18,buffalo20} and to set constraints on the cluster geometry \citep[see e.g., ][]{morandi11,morandi12,sereno18}.

The paper is organized as follows: in Sect.~\ref{sec:adv} we list the expected advantages of cluster potential over cluster HE mass in cosmological applications. We then outline the methods to derive both quantities in Sect.~\ref{sec:methods}. The cosmological simulation used in this study is described in  Sect.~\ref{sec:simu}, and analyzed in Sect.~\ref{sec:syst}. The results are discussed in Sect.~\ref{sec:disc} and lead us to conclude in Sect.~\ref{sec:conc}. As we aim at comparing the systematics (scatter and bias) inherent to the HE mass and to the reconstructed potential without accounting for the additional bias due to projecting effect, all profiles shown in this paper are 3D profiles.

\section {Expected advantages of cluster potentials}\label{sec:adv}
The gravitational potential is expected to have a number of advantages with respect to the mass in cosmological studies \citep[see also][]{lau11}:\begin{itemize}
\item{}The gravitational potential is directly related to cluster observables, such as the X-ray emission of the ICM gas and the gravitational lensing signals, without requiring the integration over an unknown volume within unknown boundaries. Therefore, the gravitational potential provides a more direct connection to cluster observables, than the cluster mass. \\

\item{} We are more flexible with the choice of the region on which we measure the potential, while the cluster mass needs to be computed from the cluster center up to a given radius. For instance, baryonic physics is complicated in the cluster center and our simple assumptions may not hold there. The strength in using the potential is that we can exclude this region from our analysis. This has been shown for instance in \citet{lau11}, where the cluster potential has been estimated from simple assumptions for different regions in the cluster: once the central region excluded, the potential estimator is a good representation of the true potential. This is not true for mass where every mass within the cluster boundary, including the core, is accounted for. \\

\item{} The gravitational potential can be derived from the mass density distribution by integrating the Poisson equation.  This implies that the  potential is smoother and  more spherical than the mass density distribution. Indeed, expressing this relation in Fourier space, we can see that the small scales fluctuations (large $k$) in the potential are suppressed ($\hat{\Phi}\propto \hat{\rho}/k^2$) compared to the mass density distribution. From the superposition principle, this also implies that large-scale modes contribute more strongly to the potential than to the density.  As the potential is less affected by the assumption of a simple cluster geometry, this may reduce the scatter in scaling relations due to any non-sphericity or clumpiness in the matter distribution. In addition, the potential is only defined up to an additive constant and corresponds to the second derivative of the density. This implies that our choice to set the potential to zero at a given radius, usually chosen to be far from the cluster center, does not influence the density distribution. The choice of the boundary conditions, required to integrate the Poisson equation, should therefore have less impacts on the potential than it as on the mass, as we derive the latter by integrating the density distribution over a volume of unknown shape (see first bullet above).\\

\item{} The gravitational potential is set early during the formation of dark matter halos \citep{vandenbosch14}. As opposed to halo mass, the potential does not suffer from pseudo-evolution, in which the mass of any given DM halo, defined as a function of the background density of the universe, increases artificially simply due to the change in the background density \citep{diemer13}. \\

\end{itemize}
For these reasons, we argue that the potential should be a quantity easier to handle and to link to theoretical predictions than the cluster mass. In addition, it has been shown in  \citet[][]{angrick15} that it is possible to constrain cosmology by studying the statistics of the cluster population characterized by their gravitational potentials.

In order to advance the use of potential over mass, we examine here how well cluster masses and potentials can be constrained from observations. We will now describe the methods used in this work to  estimate these quantities. 

\section {Methods for cluster mass and  potential estimates}\label{sec:methods}

In this section, we describe the methods used to estimate the cluster HE mass and the potential from the multi-wavelength observations of the ICM gas.

\subsection {Methods to estimate cluster masses}\label{sec:mass_method}
Under the hydrostatic equilibrium (HE) assumption, the pressure and gas density profiles recovered from X-ray or/and tSZ observations can be converted into the cluster gravitational potential:
\begin{equation}\label{eq:eqHydro}
  \nabla P=-\rho \nabla\Phi\;,
\end{equation}
where $\Phi$ is the 3D gravitational potential, while $P$ and $\rho$ are respectively the thermal pressure and density of the ICM gas. Assuming spherical symmetry, the HE mass of a cluster can then be estimated as:
\begin{equation}\label{eq:he_mass}
  M_{\rm HE}(<r)=-\frac{r^2}{G\rho(r)} \frac{dP(r)}{dr}\;,
\end{equation}
where $r$ is the 3D concentric radius. Typically HE masses are underestimated compared to the true mass due to the presence of non-thermal pressure in the clusters \cite[e.g.,][]{lau09} and to the steady-state assumptions \citep[which enters the HE equation by setting $dv/dt$=0 in the Euler equation, see e.g.,][]{suto13,lau13}. While the non-thermal pressure support accounts for the turbulent and bulk motions, the steady-state assumptions neglect the acceleration of gas that provides extra support against gravity. In dynamically relaxed clusters, the non-thermal pressure support is expected to reach up to 35\% in the cluster outskirts due to accretion of matter \citep[e.g., ][]{nagai07,rasia12,vazza11,nelson14b,biffi16}. In contrast, the deviations due to the steady-state assumptions are expected to be small (of the level of 3\%) for these clusters \citep{nelson14}. 
The hydrostatic mass bias can be quantified as: 
\begin{equation}
1-\frac{M_{\text{HE}}(<r)}{{M}(<r)}, 
\end{equation}
where $\text{M}_{\text{HE}}(<r)$ is the hydrostatic cluster mass and $\text{M}(<r)$ the true mass contained within a sphere of radius $r$. Given that the weak lensing mass does not require knowledge on the dynamical state of the matter, it is often used as a proxy of the true cluster mass. This however neglects the projection effects, which bias low the weak lensing masses by $\sim$ 5-10~\% \citep[e.g., ][]{meneghetti10,rasia12,henson16}. 

Alternatively, scaling relations can be used to estimate cluster masses from observations of the ICM gas \citep[using quantities directly observable like the X-ray luminosity or temperature, e.g., ][]{giodini13,kaiser86}. These relations are embedded in a substantial scatter \citep[see also][]{mantz10}, which can be reduced  by calibrating the HE mass with the gravitational lensing mass \citep[as in][]{linden14,applegate16} or by excluding the emission of the cluster core from the total luminosity \citep[e.g.,][]{mantz17}. 

\subsection {Methods to estimate cluster potentials}\label{sec:pot_method}
Here we outline the method we use in this work to reconstruct the gravitational potential of galaxy clusters. Other approaches exist in the literature \citep[see e.g.][]{lau11,gifford14}, but our method has a higher potentiality to reach lower bias and variance in the reconstructed potential, as we will discuss in Sect.~\ref{sec:disc}. The cluster potential can be reconstructed from the X-ray emission of the ICM gas \citep[][]{konrad13,tchernin15} and the tSZ signal \citep[][]{majer16} based on the following assumptions \citep[see also][]{tchernin18}:

\begin{enumerate}
\item{} The plasma pressure follows a polytropic relation, with a constant polytropic exponent $\Gamma$: 
\begin{equation}\label{eq:eqPoly}
  \frac{P(r)}{P_0}=\left(\frac{\rho(r)}{\rho_0}\right)^{\Gamma}\;,
\end{equation}
where $r$ parametrizes the distance from the cluster center (for instance, the 3D concentric radius, in case of assumed spherical symmetry). The suffix $0$ corresponds to the value of the pressure and of the gas density at an arbitrary fiducial radius $r_0$.  We assume in this study that $\Gamma$ is constant, but $\Gamma$ is usually observed to vary over the cluster range \citep[e.g.,][]{shaw10} and to have a value close to $\sim$1.2 \citep[e.g., ][]{tozzi01,capelo12,eckert13,shi16}. Whereas the origin of this value is still unclear, numerical simulations show that at the cluster center, the value of $\Gamma$ tends to the adiabatic value (5/3)  \citep[due to the formation history of the cluster,][]{rabold17}. In the present study, we derive the value of $\Gamma$ from a fit of the simulated density and pressure profiles over a region that excludes both the central region and the cluster outskirts. We will discuss the effect of this assumption on the reconstruction of the potential in Sect.~\ref{sec:statAn_Potpoly}. \\

\item{} The plasma is in HE with the potential of the cluster (Eq.~[\ref{eq:eqHydro}]). 
As discussed in Sect.~\ref{sec:mass_method},
this assumption is not expected to be valid over the entire cluster range. For instance, the contribution of the non-thermal pressure component increases with the distance to the cluster center due to accretion of matter in the outskirts \citep[e.g., ][]{vazza11,reiprich13,nelson14b,vazza18}. We will examine the bias coming from this assumption on the HE mass and potential in Sect.~\ref{sec:statAn_HE}. \\
\item{} The cluster has a simple geometry. In the present study, we assume spherical symmetry in order to compare the bias in the reconstructed potential with the one in the HE mass, estimated from Eq.~[\ref{eq:he_mass}]. The cluster potential can however be derived for more realistic shapes. We can for instance reconstruct the potential of spheroidal clusters \citep[whose axes ratios satisfy 1:1:a with a$>$1, as depicted in Fig.~\ref{fig:sketch}; see also][]{majer16}. We will examine the systematics of the potential reconstruction arising from the triaxiality of the clusters in our upcoming paper (Tchernin et al., in prep.). \\
\item{} For the potential reconstruction from X-rays, we assume that the X-ray emission is dominated by thermal Bremsstrahlung. This assumption is not valid for low temperature clusters ($<$ 10$^7$K) or for clusters with many substructures, due to the  contribution of emission lines to the X-ray emission \citep[see][]{sarazin88,kaastra08}. We will test the effects of this assumption on the  potential reconstructed from the X-ray signal in Sect.~\ref{sec:statAn_xraybrem}.

\end{enumerate}
To reconstruct the potential from the above assumptions, we proceed as follows: we first deproject\footnote{In the present study, we only consider simulated quantities that are already in 3D, so this deprojection step is not required to recover the 3D potential.} the observed 2D quantity for an assumed cluster morphology \citep[with for instance the Richardson-Lucy deprojection method,][]{lucy74,lucy94}. Then  we convert the resulting 3D profile (which can be the pressure profile ($P_{tSZ}$) for the observations of the tSZ signal; or the X-ray emissivity ($j_x$)) into the 3D gravitational potential by computing

\begin{equation}\label{eq:recX}
\Phi\propto [j_x]^{\eta}, \,\,\text{ with }\,\, \eta = \frac{2\Gamma-1}{3+\Gamma}, 
\end{equation}
for a reconstruction from X-rays, and
\begin{equation}\label{eq:recSZ}
\Phi\propto [P_{tSZ}]^{\eta}, \,\,\text{ with }\,\, \eta = \frac{\Gamma-1}{\Gamma},
\end{equation}
for a reconstruction from the tSZ effect. We invite the interested reader to look at the  papers by \citet{konrad13} and \citet{majer16} for details on the intermediate steps.

This 3D reconstructed potential can then be projected into 2D and combined with lensing constraints  \citep[as performed in][]{huber19}. The resulting joint 2D potential is tightly constrained on a wide radial range and could ultimately be used in cosmological studies as a good alternative to the cluster mass \citep[see e.g.,][and Sect.~\ref{sec:adv}]{angrick15}. We will next compare the systematics of the reconstructed cluster potential with those of the cluster mass, both estimated from ICM observations. 

\section {The simulation}\label{sec:simu}
We use the {\em Omega500} simulation \citep{nelson14} which is a cosmological hydrodynamic simulation of galaxy clusters that assumes a $\Lambda$CDM cosmology (H$_0 =70$ km s$^{-1}$ Mpc$^{-1}$, $\Omega_{\text{M}}=0.27$, $\Omega_{\Lambda}=0.73$). The {\em Omega500} simulations are performed using the Adaptive Refinement Tree (ART) N-body+gas-dynamics code \citep{kravtsov99, rudd08}. 
The box size of the simulation is $500 h^{-1}$Mpc on the side, with dark matter particle masses of $1.1\times 10^9 h^{-1}M_\odot$ within $5\times R_{500c}$ for the 85 resolved DM halos with $M_{500c} \geq 2.9\times 10^{14} h^{-1} M_\odot$. The highest spatial resolution is $3.8 h^{-1}$kpc.  For this paper, we only consider the non-radiative run of the simulation which does not contain cooling and feedback physics that dominates the cluster cores. 


\section {Systematics of the potential and HE mass reconstructions}\label{sec:syst}
We compare here the systematics of the reconstructed potentials with the ones of the HE masses for the clusters of the {\em Omega500} simulation. We assume spherical symmetry and present our results as 3D profiles. 


This section is structured as follows: in Sect.~\ref{sec:testcases} we illustrate the method to reconstruct clusters potential through a detailed study of two simulated clusters in extreme dynamical states: a relaxed (CL135) and a merging (CL77) cluster. In Sect.~\ref{sec:statAn} we carry out a statistical analysis of the 85 simulated clusters of the {\em Omega500} simulation: we investigate how the systematics of the HE mass and the potential compare for clusters in different dynamical states  (in Sect.~\ref{sec:statAn_dynstate}) and we study how each individual assumption listed in Sect.~\ref{sec:methods} affects the reconstructed potentials and HE masses. In particular, we test the effect of the HE assumption in Sect.~\ref{sec:statAn_HE}, the effect of the assumed polytropic stratification relation on the reconstructed potential in Sect.~\ref{sec:statAn_Potpoly}, and the effect of the assumed Bremsstrahlung dominated X-ray emissivity in Sect.~\ref{sec:statAn_xraybrem}.


In addition, we examine the effect of the presence of substructures. The latter should be removed before proceeding to the analysis because their thermodynamic properties are not representative of the ones of the ICM gas. However this technical step should be performed carefully because it can modify the morphology of clusters and introduce additional bias. We consider here two different substructure removal techniques that are widely used both in the domain of observations and simulations. We will investigate their effects on the HE mass and potential reconstruction in Sect.~\ref{sec:statAn_subs}. These two approaches are:
\begin{itemize}
    \item \textit{Method 1}: In spherical shells, we report the median rather than the mean value of a density distribution. As shown in \citet{zhuravleva13}, the mean value is biased high in the presence of density inhomogeneities while the median value can be used as a proxy for the ICM bulk density. This method has been used in many analyses \citep[such as][]{eckert18, tchernin16}. We note however, that while this substructure removal approach can be successfully applied to systems that are close to spherical, removing substructures in spherical shells for an ellipsoidal cluster could actually bias low the resulting profiles. This is illustrated in Fig.~\ref{fig:triaxial_median}, which shows the relative residuals between the \textit{mean} and \textit{median} values of density distributions extracted in spherical shells, for cluster shapes of increasing ellipticity (we consider here $\epsilon=$0, 0.3 and 0.6). As there are no substructures in these clusters, the mean and the median values in each shell should be similar. However, as we can see the \textit{median} profile steepens with respect to the \textit{mean} profile for the cluster with the largest ellipticity\footnote{The small amplitude ($<1\%$) wiggles around the 0 value  are due to  numerical artifacts introduced by the software used to create these clusters.}. This illustrates that the median approach needs to be applied carefully to non-spherical
    clusters. We will return to this point in Sect.~\ref{sec:disc}.\\

    \item \textit{Method 2}: In spherical shells, we exclude the pixels with density values larger than the threshold $\rho_{thr}$ = \textit{m} + 3.5$\sigma$, where  \textit{m} is the median density value and $\sigma$, the standard deviation in the shell. This approach is being used in \citet[][]{lau15} for instance. This threshold value has the property to clean the cluster from its substructures without forcing it to be artificially spherical (as a stronger cut would remove structures that are actually belonging to the cluster’s body). We illustrate this effect in Fig.~\ref{fig:Xsigma_map}, where we can see how this threshold affects the shape of a relaxed cluster (CL135), projected along the x-axis. We test here two different values for the threshold X$\sigma$: X=2 and X=3.5. As we can see, the stronger cut (X=2) artificially makes the inner part of the cluster CL135 more spherical, which is not a wanted feature.

\end{itemize} 

For completeness, we also show in Fig.~\ref{fig:Xsigma_profile} the relative residuals between these two methods, applied to the cluster CL135, for different values of the threshold X=1, 2, 3.5 and 4. For this cluster, the median method (\textit{Method 1}) seems to be similar to using \textit{Method 2} with a cut of 2$\sigma$, which as we saw in Fig.~\ref{fig:Xsigma_map}, makes the cluster more spherical.
It is not an easy task to remove substructures without modifying the shape of the cluster and affecting the resulting profiles. In the present study, we assume spherical symmetry, so the effect produced by the substructure removal technique is expected to be small with respect to the strong assumption we are making. However, the choice of the substructure removal technique  needs to be carefully studied once we include the triaxiality of the clusters in our reconstructions. This effect and the resulting systematics will be analyzed in detail in our upcoming paper (Tchernin et al., in prep.).

\subsection{Test cases of a relaxed and a dynamically active cluster}\label{sec:testcases}

\begin{figure}
\begin{subfigure}{.5\textwidth}
  \centering
  \includegraphics[width=\columnwidth,angle=0]{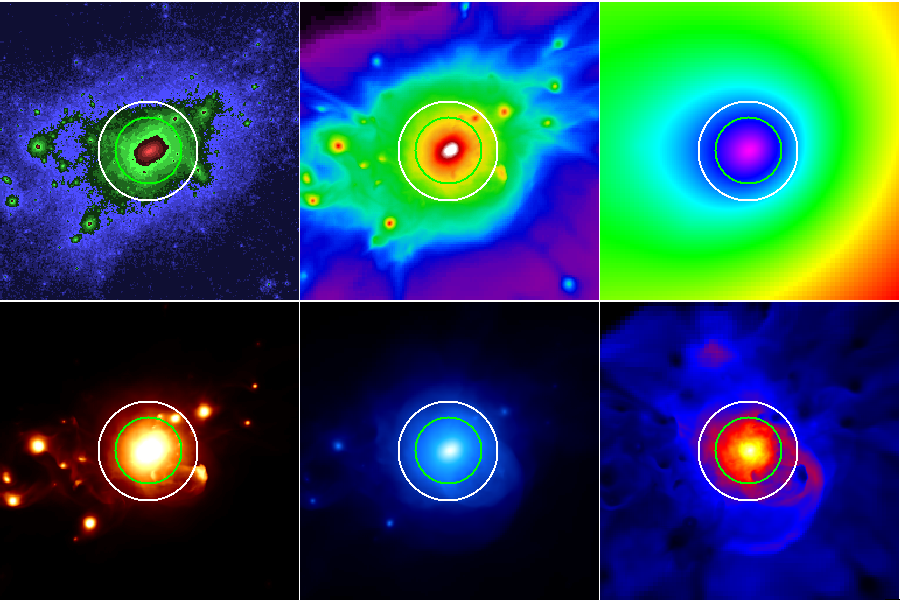}
  \caption{}
   \label{fig:snapshotcl135}
\end{subfigure}
\begin{subfigure}{.5\textwidth}
  \centering
  \includegraphics[width=\columnwidth,angle=0]{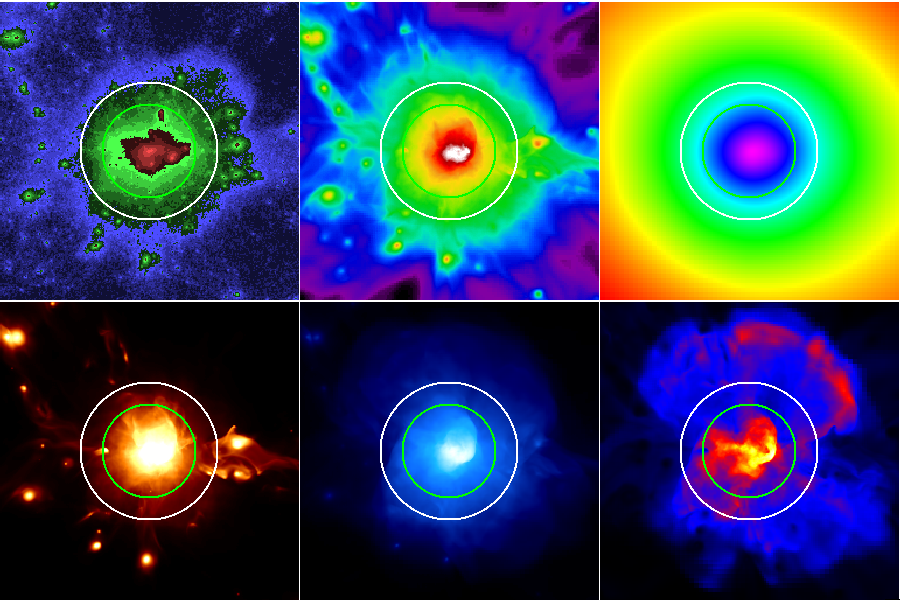}
  \caption{}
 \label{fig:snapshotcl77}
\end{subfigure}
\caption{Component decomposition and ICM gas emission of: (a) the relaxed cluster CL135, and (b) the merging cluster CL77 (see Sect.~\ref{sec:testcases}). Top panels, from left to right: the projected dark matter mass; the projected gas mass; and the projected cluster gravitational potential. Bottom panels, from left to right: the projected X-ray emission; the projected tSZ effect; and the mass-weighted temperature.  The two circles represent $\text{R}_{500}$ (in green) and $\text{R}_{200}$ (in white).}
\label{fig:snapshot}
\end{figure}

We illustrate here the potential and HE mass reconstruction methods on two clusters in extreme dynamical states: the relaxed cluster CL135 and the dynamically active cluster CL77. These clusters have been selected based on their mass accretion histories and their X-ray morphologies. The growth of CL135 is mainly driven by smooth accretion, having no recent major merger (defined as mass ratio between merging halos $>$1:6) in the past Gyr, while CL77 has experienced one in the past Gyr.   The morphological classification is based on the observation-motivated values of the symmetry, peakiness, and alignment (\textit{SPA}) measurements of the X-ray isophotes of the clusters \citep[][]{mantz15}. We generated mock {\em Chandra} maps of these clusters and apply the \textit{SPA} criterion to them \citep[see][]{shi16b}. A view of diverse components of these two clusters is given in Fig.~\ref{fig:snapshot}.

\subsubsection{Method and Results}
We use the method outlined in Sect.~\ref{sec:pot_method} to reconstruct the cluster potential from the X-ray and tSZ observations (hereafter the X-ray and tSZ potential), and we recover the HE mass by solving the HE equation (Eq.~[\ref{eq:he_mass}]).  

\begin{figure}[h!]
\begin{subfigure}{.5\textwidth}
  \centering
  \includegraphics[width=\columnwidth,angle=0]{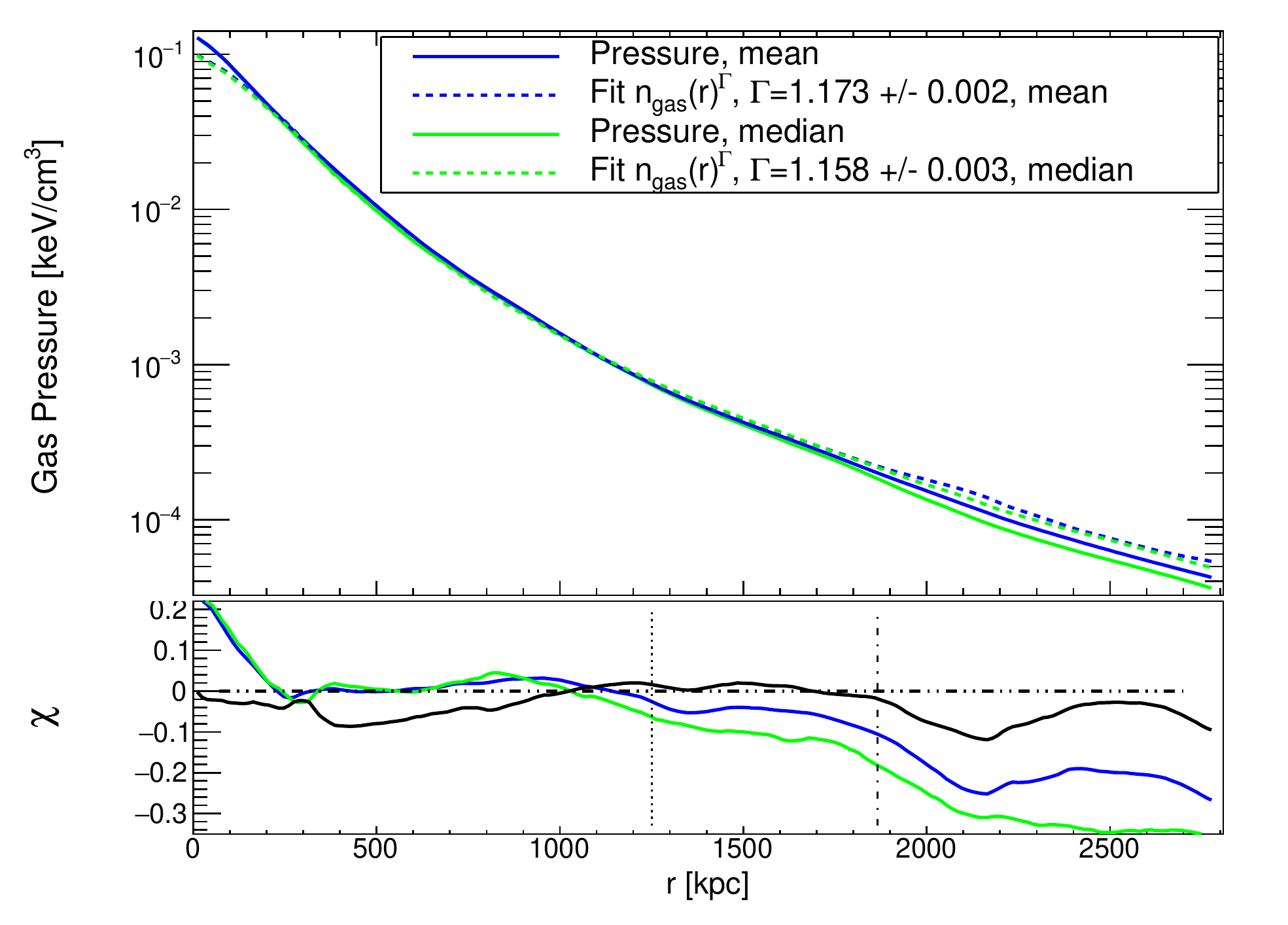}
  \caption{}
   \label{fig:polyindexcl135}
\end{subfigure}
\begin{subfigure}{.5\textwidth}
  \centering
  \includegraphics[width=\columnwidth,angle=0]{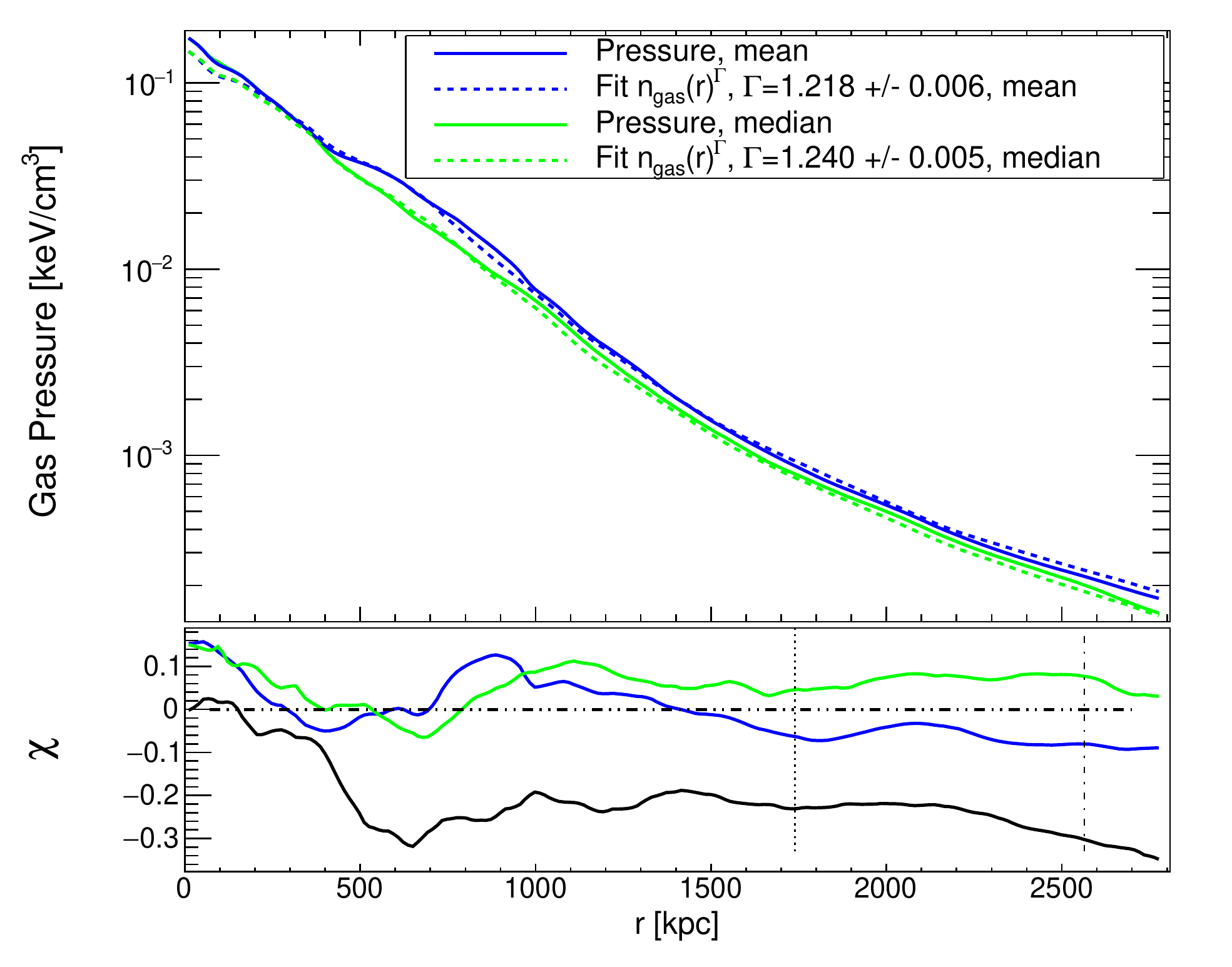}
  \caption{ }
   \label{fig:polyindexcl77}
\end{subfigure}
\caption{Fit of the polytropic stratification relation on the pressure [kev/cm$^3$] and gas density [cm$^{-3}$] profiles extracted in spherical shells.  Top panel: pressure profile (solid line) and its fitted model (dashed line) are shown in blue, for the \textit{mean} profiles and in green, for the \textit{median} profiles.  Bottom panel: the relative residuals between the pressure profile and its fitted model are shown in blue, for the \textit{mean} profiles and in green, for the \textit{median} profiles. The relative residuals between the two models for the pressure profile, computed as $((n_{\text{gas,median}}(r))^{\Gamma\text{median}}-(n_{\text{gas,mean}}(r))^{\Gamma\text{mean}})/(n_{\text{gas,median}}(r))^{\Gamma\text{median}}$ are displayed in black. The vertical lines represent $\text{R}_{500}$ and $\text{R}_{200}$; 
a) For CL135: Fit performed over radii larger than 200 kpc;
b) For CL77: Fit performed over radii larger than 350 kpc for the median profiles and over radii larger than 200 kpc for the mean profiles.}
\label{fig:polyindex} 
\end{figure}

We start by removing substructures in these clusters with the technique of the median (see the above description of \textit{Method 1}). To reconstruct the potential, we then fit the polytropic relation on the pressure and density profiles. The result of this fit is shown in Fig.~\ref{fig:polyindexcl135}, for the relaxed cluster and in Fig.~\ref{fig:polyindexcl77}, for the merging cluster. We show both the result of the fit of the \textit{median}   and  \textit{mean} profiles to illustrate the effects of the substructures (for this substructure removal technique). Assuming spherical symmetry and hydrostatic equilibrium, we then reconstruct the X-ray and tSZ potential, and the HE mass of these two clusters.

\begin{figure}[h!]
\begin{subfigure}{.5\textwidth}
  \centering
  \includegraphics[width=\columnwidth,angle=0]{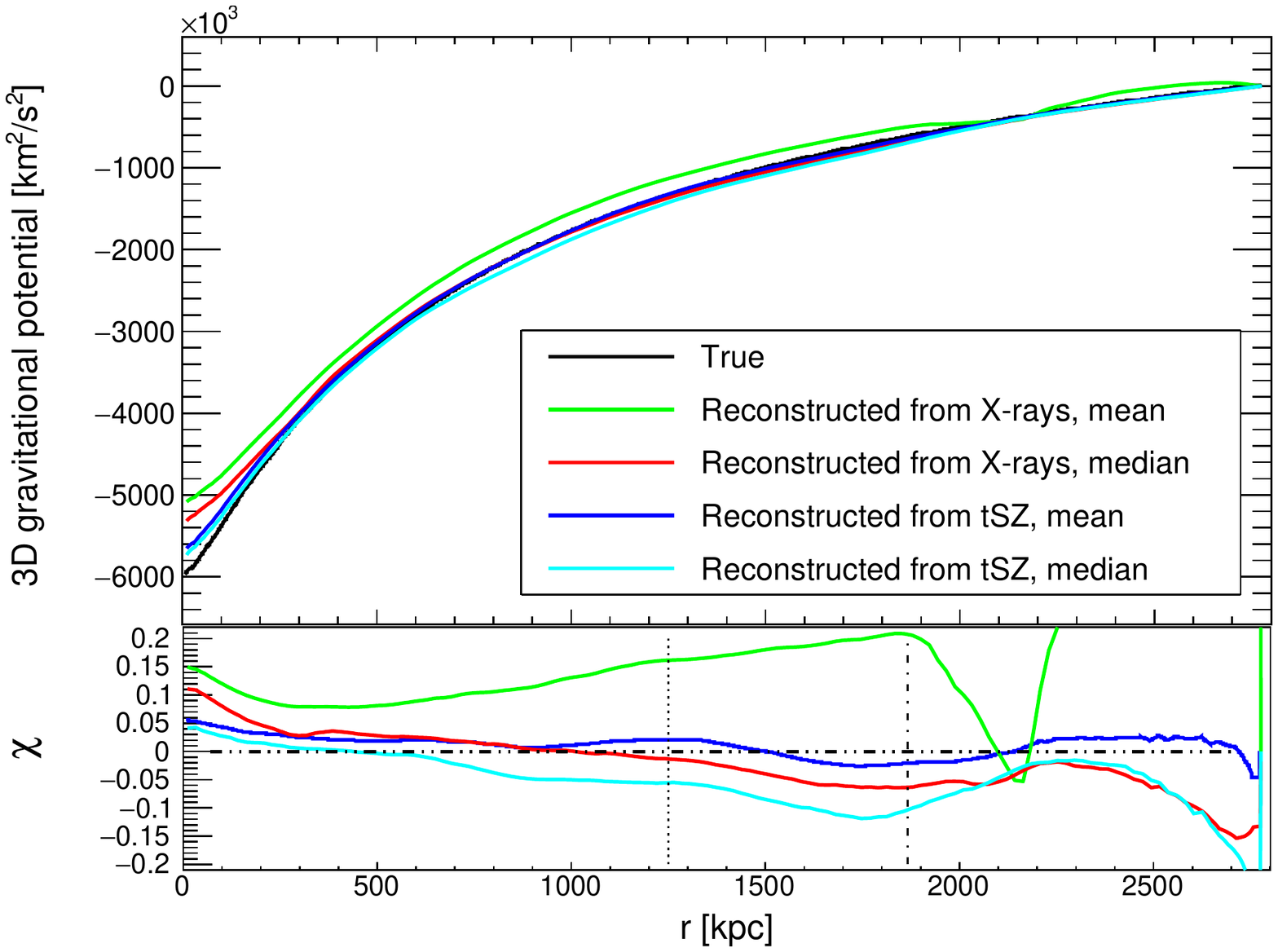}
  \caption{}
   \label{fig:potcl135}
\end{subfigure}
\begin{subfigure}{.5\textwidth}
  \centering
  \includegraphics[width=\columnwidth,angle=0]{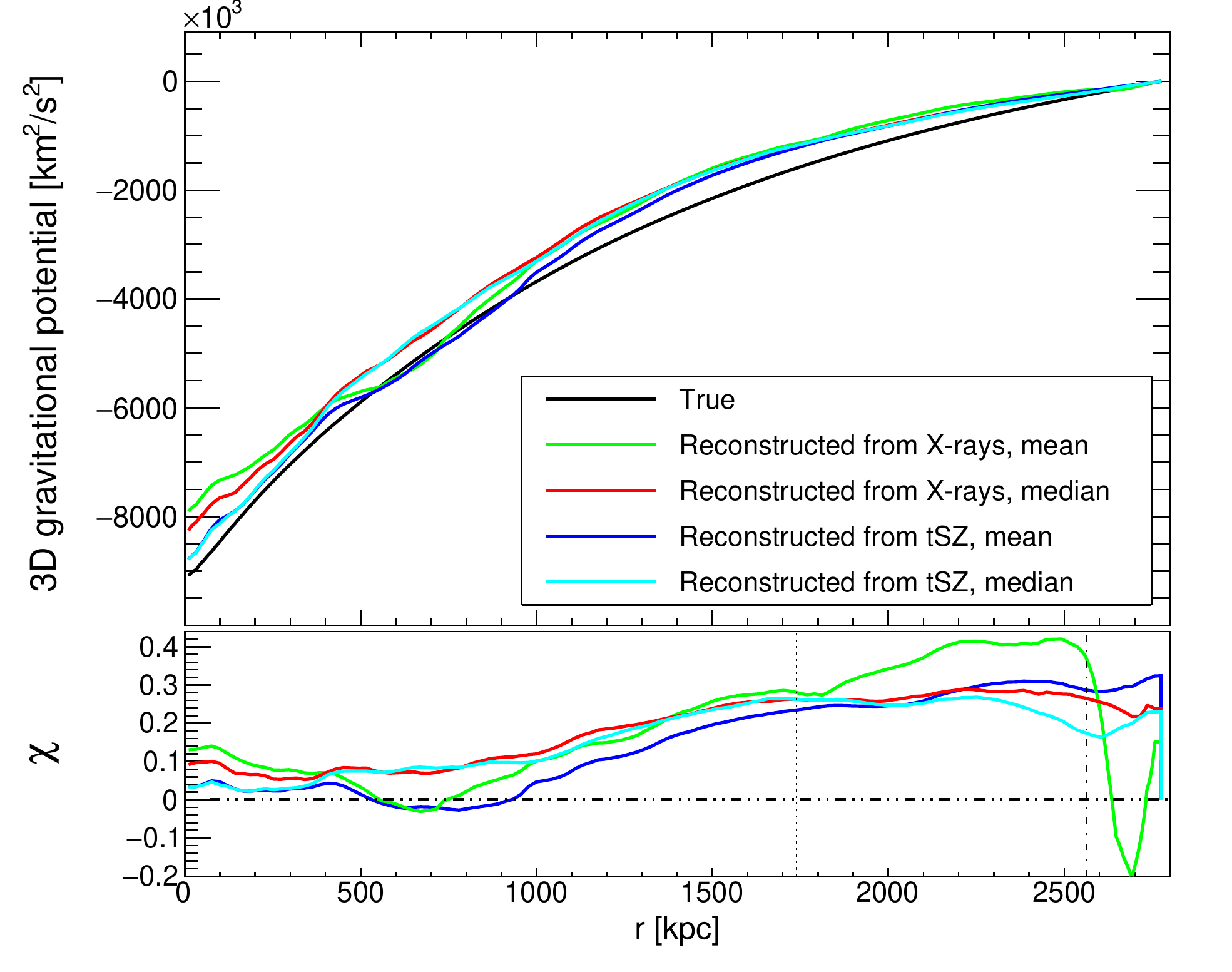}
  \caption{ }
   \label{fig:potcl77}
\end{subfigure}
\caption{Reconstructed 3D gravitational potential profiles obtained as outlined in Sect.~\ref{sec:pot_method} assuming spherical symmetry. Top panel: true potential (black curve); X-ray potential \textit{mean} (green) and \textit{median} (red); tSZ potential \textit{mean} (blue) and \textit{median} (cyan). Bottom panel: corresponding relative residuals. The vertical lines represent $\text{R}_{500}$ and $\text{R}_{200}$: a) For the relaxed cluster CL135; b) For the merging cluster CL77.}
\label{fig:pot}
\end{figure}

\begin{figure}[h!]
\begin{subfigure}{.5\textwidth}
  \centering
  \includegraphics[width=\columnwidth,angle=0]{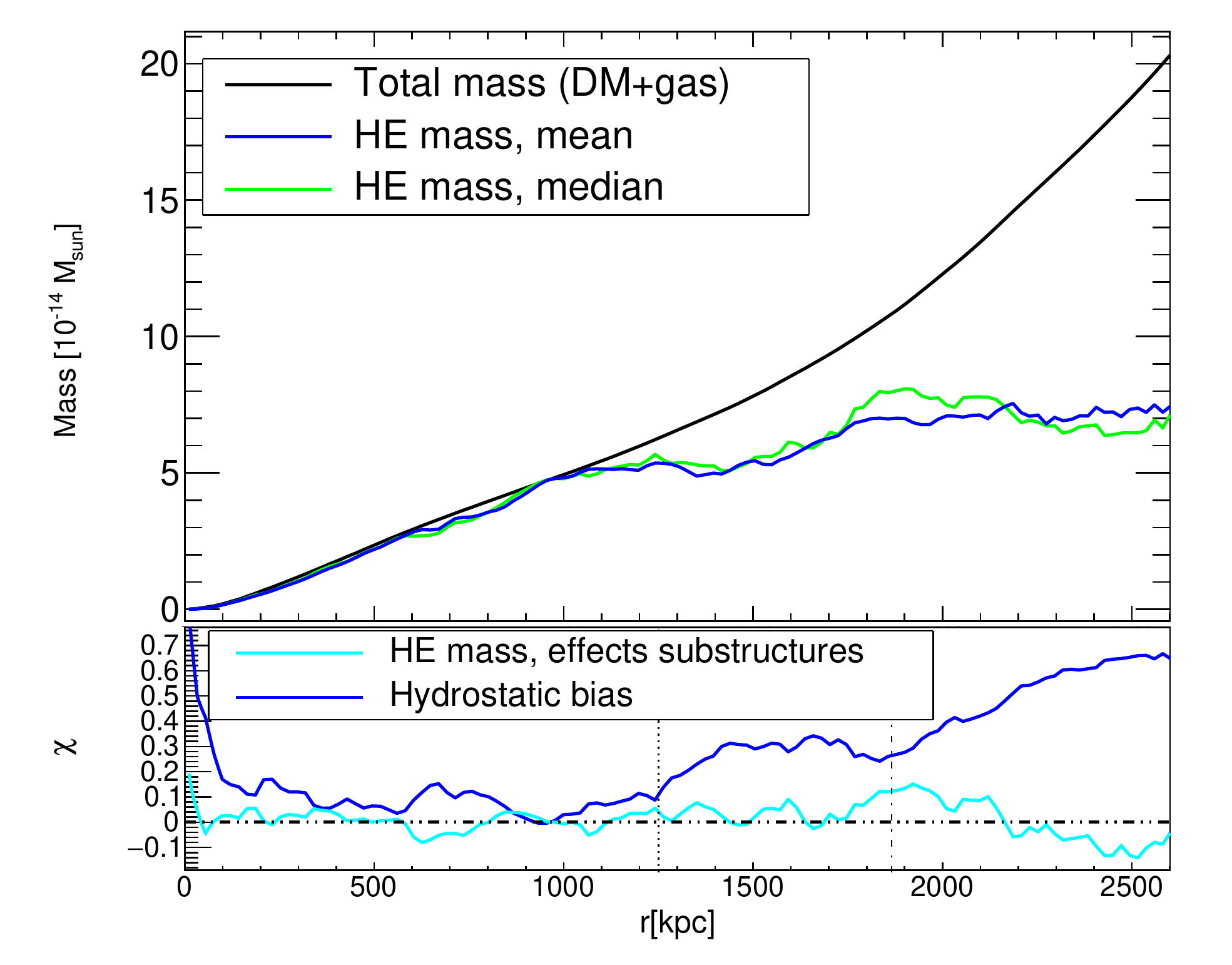}
  \caption{}
   \label{fig:MHE_Motcl135}
\end{subfigure}
\begin{subfigure}{.5\textwidth}
  \centering
  \includegraphics[width=\columnwidth,angle=0]{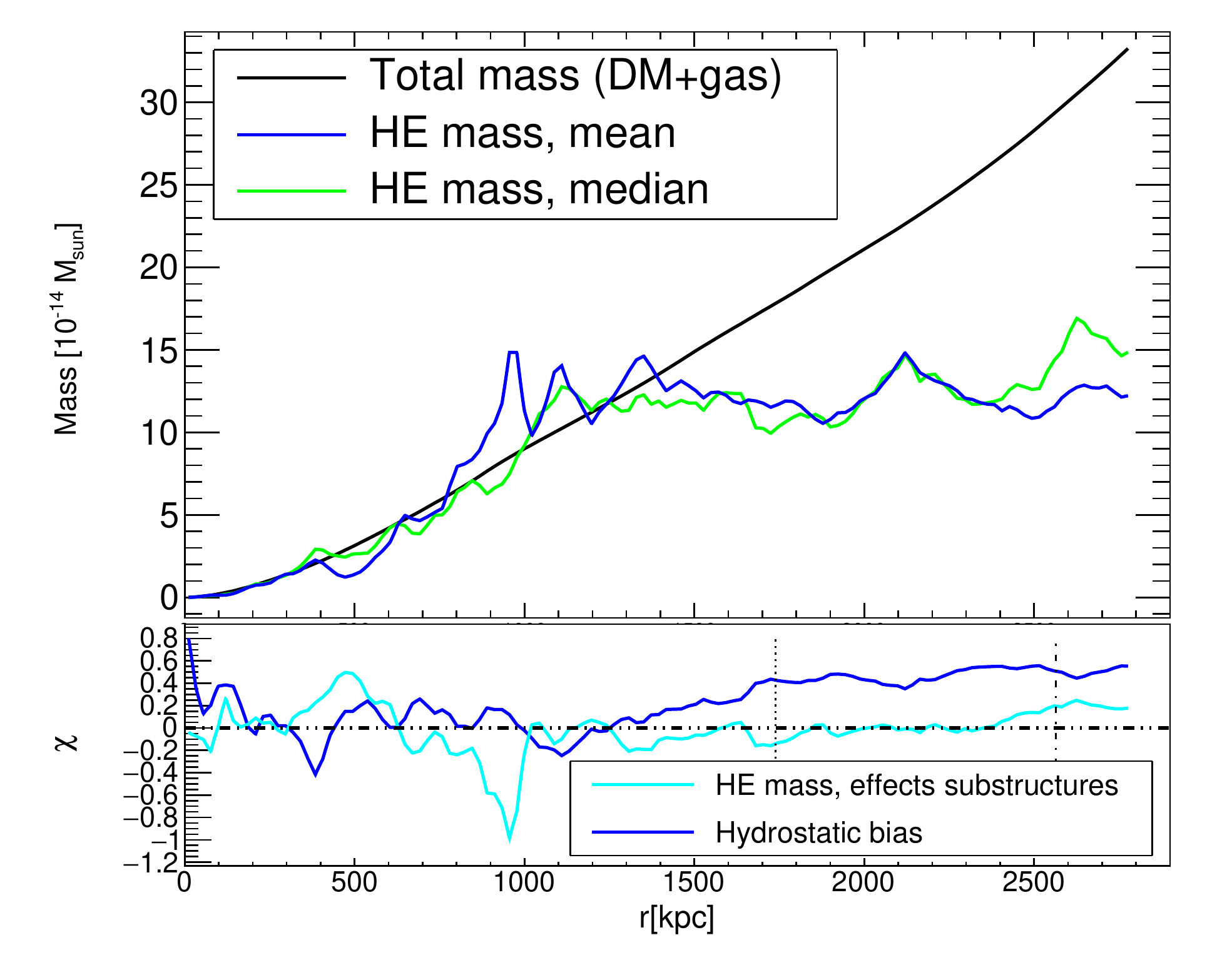}
  \caption{ }
   \label{fig:MHE_Motcl77}
\end{subfigure}
\caption{Reconstructed HE mass ($\text{M}_{\text{HE}}$) obtained assuming spherical symmetry. Top panel: total mass ($\text{M}_{\text{tot}}$, black curve), $\text{M}_{\text{HE}}$ \textit{median} (green) and $\text{M}_{\text{HE}}$ \textit{mean} (blue). Bottom panel:  relative residuals: the hydrostatic bias (computed as $(\text{M}_{\text{tot}}(r)-\text{M}_{\text{HE}_{\text{median}}}(r))/\text{M}_{\text{tot}}(r)$), is shown in black; and the effect of substructures  ($(\text{M}_{\text{HE}_{\text{median}}}(r)-\text{M}_{\text{HE}_{\text{mean}}}(r))/\text{M}_{\text{HE}_{\text{median}}}(r))$, in cyan. The vertical lines represent $\text{R}_{500}$ and $\text{R}_{200}$: a) For the relaxed cluster CL135; b) For the dynamically active cluster CL77.}
\label{fig:MHE_Mot}
\end{figure}

The reconstructed potentials are shown in Fig.~\ref{fig:potcl135} and~\ref{fig:potcl77}, for the clusters CL135 and CL77 respectively. In each figure, the reconstructed X-ray and tSZ potentials (for both mean and median profiles) are shown in the top panel; while the bottom panel displays the relative residuals between the true and reconstructed potentials. 
The HE mass estimates of these two clusters are represented in the top panel of Fig.~\ref{fig:MHE_Motcl135} and~\ref{fig:MHE_Motcl77}. The relative residuals, shown in the bottom panel, highlight the effects of substructures and the hydrostatic mass bias. We will next discuss these results.

\subsubsection{Effects of substructures}\label{sec:testcases_sub}
\begin{table*}

\begin{center}
\begin{tabular}{|c|c|c|c|}\hline
            &  at R$_{500}$ & in between &at R$_{200}$\\\hline     
$\text{M}_{\text{HE}}$, relaxed        &   10\%   & $\lesssim$35\% & 30\%     \\\hline  
$\Phi_{\text{HE}}$, relaxed          &   $\sim$3-5\%   & $\lesssim$10\% & $\lesssim$8\%     \\\hline  
$\text{M}_{\text{HE}}$, unrelaxed        &   42\%   & $\lesssim$54\% & 50\%     \\\hline  
$\Phi_{\text{HE}}$, unrelaxed          &   25\%   & $\lesssim$32\% & 30\%     \\\hline  
\end{tabular}
\caption{Relative residuals  for the reconstructed potential ($\Phi$, Fig.~\ref{fig:potcl135} and ~\ref{fig:potcl77}) and HE mass ($\text{M}_{\text{HE}}$, Fig.~\ref{fig:MHE_Motcl135} and ~\ref{fig:MHE_Motcl77}) at $\text{R}_{500}$,  $\text{R}_{200}$, and in the radial range delimited by these two radii, for the relaxed (with $\text{R}_{500}=1250$ kpc and $\text{R}_{200}=1865$ kpc) and the unrelaxed (with $\text{R}_{500}=1739$ kpc and $\text{R}_{200}=2564$ kpc) clusters.}
\label{table:HE}
\end{center}
\end{table*}
\begin{itemize}
\item\textit{On the potential reconstruction: } to  more clearly distinguish this effect from those of the other assumptions, we consider here the relaxed cluster. 
 By comparing the potential reconstructed from the median and the mean profile in Fig.~\ref{fig:potcl135}, we can conclude that the removal of substructures is crucial to reconstruct the potential from the X-ray emissivity. This is due to the fact the Bremsstrahlung radiation depends on the squared density and therefore is very sensitive to density inhomogeneities. We can also notice that the tSZ potential reconstructed with the mean pressure profile provides a better result than the one reconstructed with the median pressure profile (see Eq.~[\ref{eq:recSZ}]). This is surprising because the pressure is not expected to be significantly affected by the presence of substructures, and therefore both mean and median profiles should be similar. However, as shown in Fig.~\ref{fig:pressure}, the mean and median pressure profiles of this cluster  differ (this can be also derived from the relative residuals shown in black in the bottom panel of Fig.~\ref{fig:polyindexcl135}: both pressure models are similar in the radial range between  $\text{R}_{500}$ -  $\text{R}_{200}$, while the relative residuals of the mean (in blue) and median profiles (in green) differ). 
This may be due to the substructure removal technique, which biases the profile low if the cluster is triaxial, as we discussed in the first paragraphs of Sect.~\ref{sec:syst} and in Fig.~\ref{fig:triaxial_median}. This is an interesting aspect that we will discuss further in Sect.~\ref{sec:disc}. \\
\item\textit{On the HE mass estimate}: as shown in Fig.~\ref{fig:MHE_Motcl135}, the effect of the substructures on the HE mass is smaller than 10~\% between $\text{R}_{500}$ and $\text{R}_{200}$, which is larger than for the tSZ potential, but smaller than for the X-ray potential. This indicates that the combined effects of substructures on the pressure gradient and the density profile are relatively small for this relaxed cluster. 
\end{itemize}

\subsubsection{Potential versus HE mass reconstruction}\label{sec:testcases_disc}

We consider here the X-ray potential and the HE mass reconstructed with the median profiles, while for the tSZ potential, we report the averaged value of the mean and median tSZ potential profiles. Table~\ref{table:HE} summarizes the results of these reconstructions.  For these clusters, we observe that the overall bias is smaller for the potentials than for the HE mass. Indeed, the relative residuals of the tSZ and X-ray potentials ($\sim$3-5\%) are smaller than the ones of the HE mass ($\sim$10\%) at $\text{R}_{500}$. For the dynamically active cluster, we found that the tSZ and X-ray potentials have residuals that reach $\sim$25\% at $\text{R}_{500}$ and a corresponding hydrostatic mass bias of about 42\%. We study next the recovered profiles in more detail:

   \begin{itemize}
       \item {Relaxed cluster:}
    \\As we can see in Fig.~\ref{fig:potcl135}, the X-ray (median) and the tSZ (mean) potentials reproduce the true gravitational potential with a relative deviation of $\lesssim$5\% from 300 kpc up to $\text{R}_{200}$. 
    The discrepancy at radii smaller than 300 kpc and larger than $\text{R}_{200}$ can be understood from Fig.~\ref{fig:polyindexcl135}, which shows the result of the fit of the pressure and gas density profiles assuming a polytropic stratification relation: the polytropic index is different from the fitted value in these regions \citep[as expected from simulations][]{shaw10,rabold17}. From the mass profiles displayed in Fig.~\ref{fig:MHE_Motcl135}, we can also see that the hydrostatic mass bias is about 3 times larger than the bias in the reconstructed potential. We observe that the hydrostatic mass bias increases with the  distance from the cluster center for radii larger than $\text{R}_{500}$ and reaches the predicted value of $\sim$30-35\% at $\text{R}_{200}$ due to accretion of matter in the outskirts \citep[e.g.][see also Sect.~\ref{sec:mass_method} and the bullet 2 in Sect.~\ref{sec:pot_method}]{rasia12,vazza18}.\\
    
    \item {Merging cluster:}
    \\The assumptions being far from valid for this cluster, we expect that both the HE mass and potential reconstructions perform less well than for a relaxed cluster.
    This is indeed what we observe in Fig.~\ref{fig:potcl77}: the relative residuals for the reconstructed potential reach up to 30\% within $\text{R}_{200}$ (rather than 8\% at the same radius for the relaxed cluster).  
    The results of the HE mass estimate of the cluster CL77 are shown in Fig.~\ref{fig:MHE_Motcl77}: the hydrostatic bias amounts to 50\%  within $\text{R}_{200}$. This illustrates that even for this extreme case, the potential can be better reconstructed than the HE mass. Interestingly, as for the cluster CL135 (Fig.~\ref{fig:MHE_Motcl135}) the effects due to the presence of substructures seem to be small between  $\text{R}_{500}$ and $\text{R}_{200}$.
 \end{itemize}




\subsection {Statistical Analysis}\label{sec:statAn}
The aim of this section is to compare the systematics of the reconstructed tSZ and X-ray potentials with the ones of the reconstructed HE mass for the 85 clusters of the {\em Omega500} simulation. We proceed as follows: we compute the relative residuals between the true and reconstructed quantities, evaluate them at $\text{R}_{500}$, and display the results in histograms. The key results are shown in the body of the paper, while the additional figures can be found in the appendix. 
We report the mean and root mean square values as a measure of the bias and spread  of these distributions in Tab.~\ref{tab:syst} and in the corresponding captions.  Except if stated differently, we here use only quantities that have been corrected for the presence of substructures using the 3.5$\sigma$ approach (see \textit{Method~2} in the introduction of Sect.~\ref{sec:syst}). This analysis is performed assuming spherical symmetry.
\subsubsection {Effect of the dynamical state}\label{sec:statAn_dynstate}We start by comparing the systematics of the HE mass with the ones of the tSZ potential\footnote{Given that the X-ray potential relies on an additional assumption (bullet 4 in Sect.~\ref{sec:pot_method}), we show here only the results for the tSZ potential. We will study separately the systematics of the X-ray potential in Sect.~\ref{sec:statAn_xraybrem} (see also Tab.~\ref{tab:syst}).} for clusters in different dynamical states \citep[classified using the \textit{SPA} criterion from][]{mantz15}.  

\begin{figure}[h!]
\begin{subfigure}{.5\textwidth}
  \centering
  \includegraphics[width=\columnwidth,height=0.7\columnwidth,angle=0]{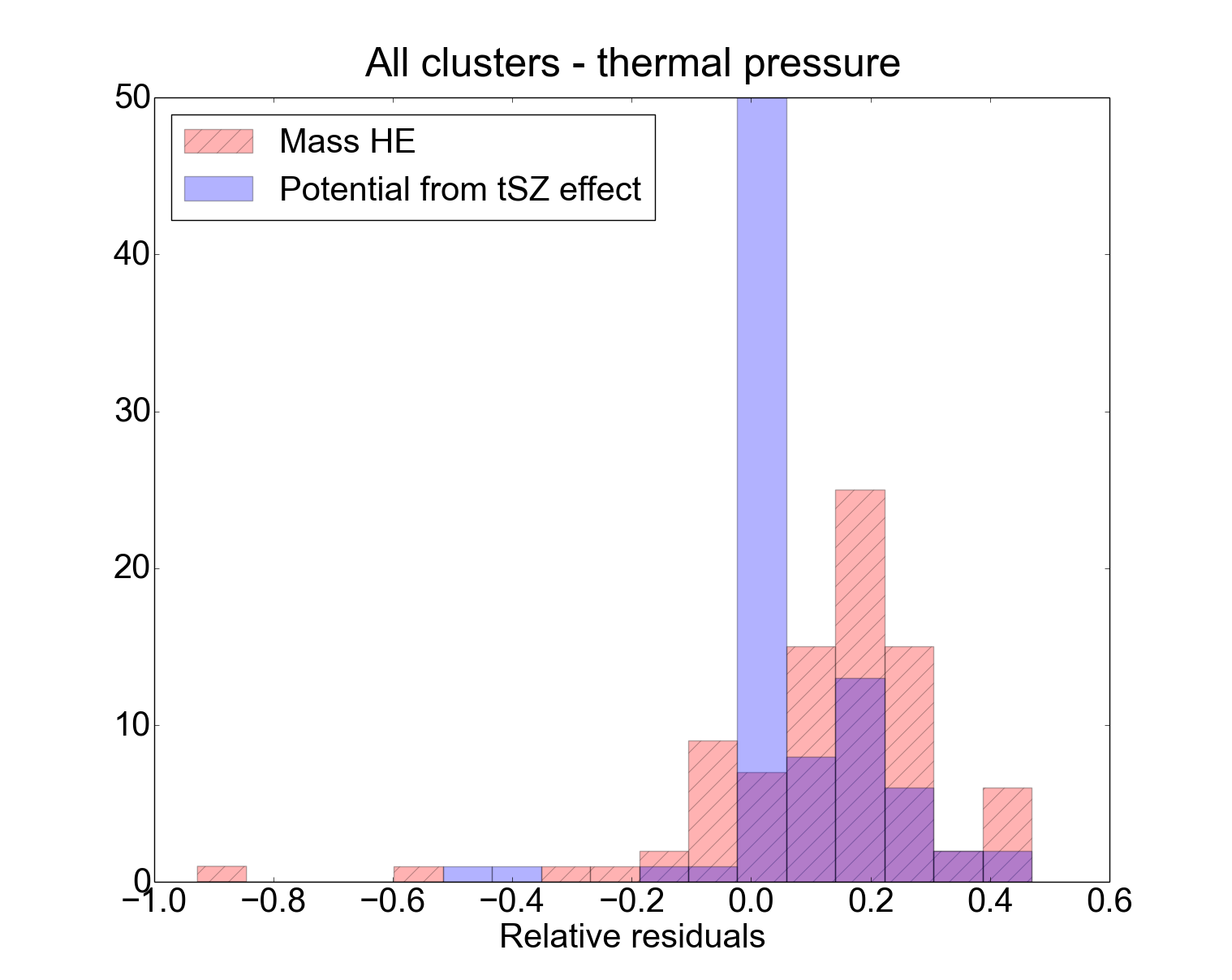}
  \caption{}
   \label{fig:statAn_dynstateAll}
\end{subfigure}
\begin{subfigure}{.5\textwidth}
  \centering
  \includegraphics[width=\columnwidth,height=0.7\columnwidth,angle=0]{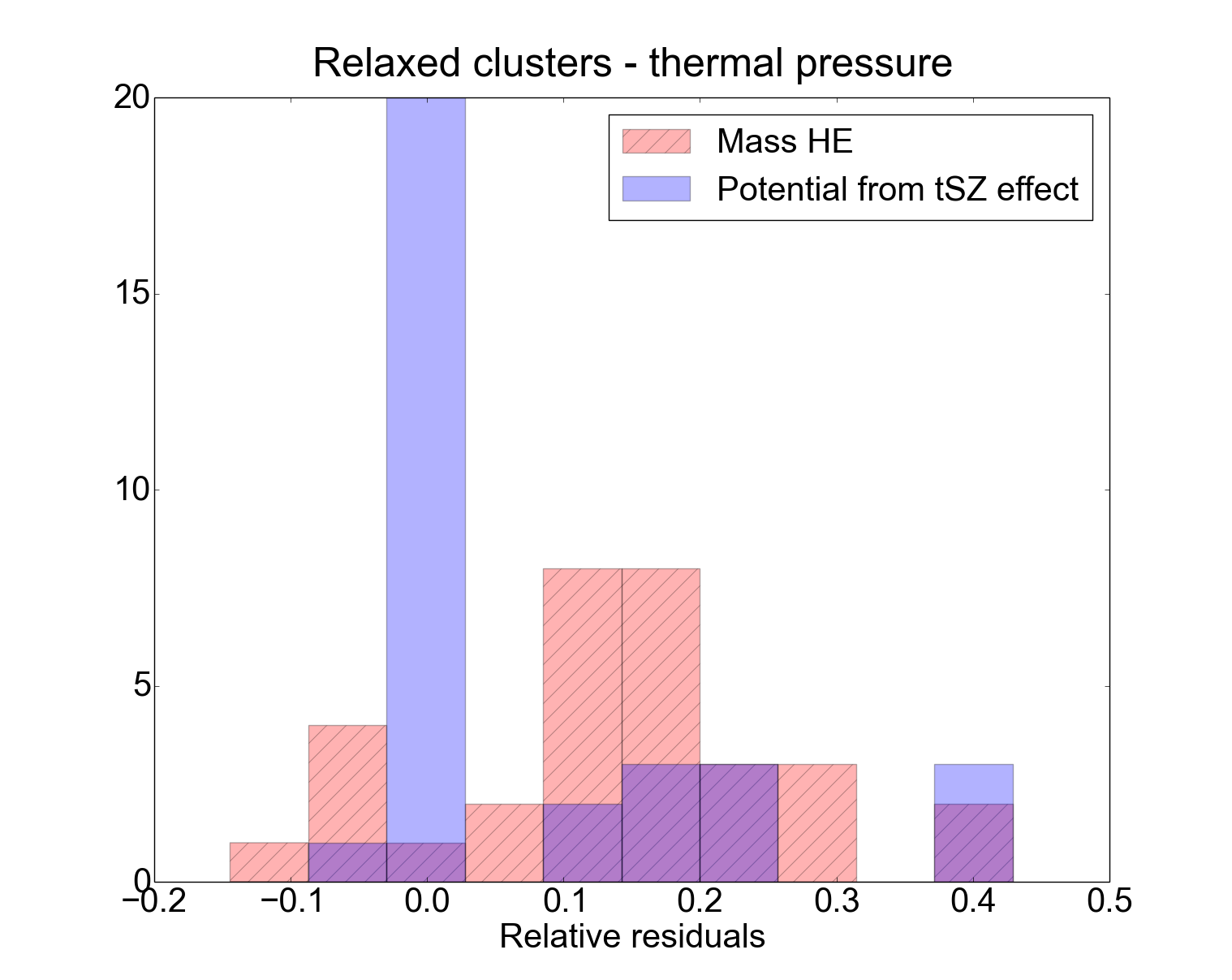}
  \caption{ }
   \label{fig:statAn_dynstateRel}
\end{subfigure}
\caption{Relative residuals - comparison of the tSZ potential and HE mass for clusters in different dynamical states, expressed as (\textit{bias}, \textit{scatter}) and evaluated at $\text{R}_{500}$ for: a) the entire sample (85 clusters), $M$=(0.13, 0.24); $\Phi$=(0.06, 0.15); b) the relaxed clusters sample (32 clusters), $M$=(0.14, 0.19); $\Phi$=(0.08,0.16).}
\label{fig:statAn_dynstate}
\end{figure}

We compute the relative residuals for the tSZ potential and the HE mass, reconstructed  using the thermal pressure.
The resulting histograms are shown in Fig.~\ref{fig:statAn_dynstateAll} for the entire cluster sample (85 clusters) and in Fig.~\ref{fig:statAn_dynstateRel} for the relaxed cluster sample (32 clusters). 
As we can see, the reconstructed potential has a smaller bias and scatter than the reconstructed HE mass, both for the relaxed clusters and for the entire cluster sample. Indeed, our results for the relaxed cluster sample return 
 a bias of 14\% (8\%) and a scatter of 19\%(16\%) for the HE mass (for the tSZ potential). The trend is similar for the entire cluster sample: with a bias of 13\% (6\%) and a scatter of 24\%(15\%) for the HE mass (tSZ potential). We will return to these results in Sect.~\ref{sec:disc}.

\subsubsection {Effect of the HE assumption}\label{sec:statAn_HE}
In this section, we are interested in quantifying the effect of the HE assumption on the reconstructed potentials and on the HE mass. We therefore reconstruct these quantities using  both  the  total  and  the thermal  pressure,  and  compare their  respective  relative residuals.  Given  that  all  the  other quantities are held fixed, any difference between these two reconstructions should be due to deviations from the HE assumption. In the {\em Omega500}  simulation, the bulk and turbulent motions of the gas  are  the main  source  of  non-thermal pressure support. However the contribution of the non-steady state terms (due to gas accelerations) is not accounted for. We consider here a sample of relaxed  clusters as the  deviations from  the  HE  assumptions  due  to  these non-steady state terms  are small for relaxed clusters, and are therefore not affecting our analysis.

The three figures displayed in Fig.~\ref{fig:statAn_HEEffect} illustrate the effect of the HE assumption on the HE mass, the X-ray and the tSZ potential. By using the total pressure instead of the thermal pressure, we observe a similar trend for the HE mass (with a bias that reduces from 14\% to 3\%) and for the X-ray potential (with a bias that reduces from 15\% to 2\%). But the tSZ potential is already reconstructed quite well with the thermal pressure (with a bias of only 8\%). 
It is interesting to see that the HE assumption seems to affect  the tSZ and the X-ray potentials differently. This discrepancy could be a hint that these two probes explore ICM components of different origin. To investigate this hypothesis, we need to account for the other effects that also enter in these reconstructions and which affect these two potential reconstructions differently (such as the substructure removal method, and the additional assumption on the X-ray emissivity\footnote{This two effects will be investigated in Sect.~\ref{sec:statAn_subs} and~\ref{sec:statAn_xraybrem}, respectively.}). We will return to this point in Sect.~\ref{sec:disc}.


\subsubsection {Effect of the polytropic stratification relation}\label{sec:statAn_Potpoly}
We saw in Sect.~\ref{sec:pot_method} that our approach to reconstruct the X-ray and tSZ potentials relies on the assumption that the density and pressure follow a polytropic relation of constant polytropic index $\Gamma$ (see Eq.~[\ref{eq:recX}] and [\ref{eq:recSZ}]).  

To evaluate how uncertainties on $\Gamma$ ($\Delta\Gamma$) affect the spread of the distributions of reconstructed potentials, we derive them for different values of $\Delta\Gamma$. We consider first the value of $\Delta\Gamma$ returned by the fit, which averaged over the entire cluster sample, returned the value of $\Gamma_{average}$=1.175 and $\Delta\Gamma_{average}$=0.003. We reconstruct the tSZ potential and observe that the spread in the distribution increases from 15\% to 16\%.  As we can attest, the reconstructed potential is almost independent from these variations in $\Gamma$.  This however is not surprising given that for these simulated clusters, the error on  $\Gamma$ is of the order of a few per-mille.
However, typical observations return percent level errors for $\Gamma$  \citep[e.g.][1.21+/- 0.03]{eckert13}, \citep[][1.17+/-0.01]{girard20}. To illustrate how such uncertainties would propagate through the reconstruction and affect the spread of the distribution of reconstructed potentials, we repeat this exercise for a $\Delta\Gamma$ 10 times larger (corresponding to $\Delta\Gamma_{average}$=0.03). This time, the spread of the distribution becomes more significant and reaches 19\%, which is still smaller than the scatter in the HE mass (24\%, see Tab.~\ref{tab:syst}).
Increasing the error on $\Delta\Gamma_{average}$ to 0.06, we observe that the scatter in the reconstructed tSZ potential (23\%) becomes comparable to that in the HE mass.

\subsubsection {Effect of the presence of emission lines in the X-ray emissivity}\label{sec:statAn_xraybrem}

The assumption that the X-ray emission is dominated by  bolometric Bremsstrahlung   (see bullet 4 in Sect.\ref{sec:pot_method}) allows us to reconstruct the potential using the formula given in Eq.~[\ref{eq:recX}]. However, this assumption is justified only if the contribution of emission lines to the total emissivity is negligible. This is the case if the plasma is very hot (T$>10^7$K) and if there are no substructures. Indeed, these density inhomogenieties have typically different metallicity and a lower temperature than the ICM gas.  
To investigate how any deviation from the bolometric Bremsstrahlung emissivity affects the potential reconstruction, we reconstruct the X-ray potentials of our cluster sample,  for both their true emissivity (which contains emission lines and is not bolometric) and for an assumed bolometric Bremsstrahlung emissivity (hereafter \textit{emissivity corrected}). We use for both reconstructions only quantities obtained after substructure removal.

 
The results are shown in Fig.~\ref{fig:statAn_embolo_allcl}. For clusters in any dynamical state, we find that the \textit{emissivity corrected} potentials are distributed with a bias of 9\%, while the potentials reconstructed from their true X-ray emissivity, have a bias of 15\% (see Tab.~\ref{tab:syst}).  The spread of both distributions are, however, comparable.

We can also note that the bias and scatter of the distribution of \textit{emissivity corrected} X-ray  potentials (0.09,0.14)   are becoming similar to those of the distribution of tSZ potentials (0.06,0.15). The remaining difference between these two reconstructions could be due to their different sensitivity to the presence of substructures (and therefore to the substructure removal technique); but it could also be caused by their different sensitivity to deviation from the HE assumption. In the latter case, this would mean that X-ray and tSZ observations are probing ICM components that are from different origin or in different physical states.
We study next how the presence of emission lines affects the X-ray potentials corrected for the HE assumption.
 To this end, we  reconstruct the X-ray potentials of the relaxed clusters sample using the total pressure instead of the thermal pressure. The results are shown in Fig.~\ref{fig:statAn_embolo_relcl}:
 as we also saw in Fig.~\ref{fig:statAn_HEeffectPotXR}, the bias using the true emissivity vanishes almost completely once the total pressure is taken into account (with a bias of 2\%). In contrast, the distribution of \textit{emissivity corrected}  potentials, reconstructed for the total pressure, overestimates the true potential with a bias of 6\%. This may be due to the spherical symmetry assumption, as we will discuss in Sect.~\ref{sec:disc}. Nevertheless, the fact that a bias of about 2\% is reached once we correct for deviations from the HE assumption shows that the effect due to the assumed bolometric Bremsstrahlung emissivity should not be significant for this sample of relaxed clusters.


\subsubsection {Effect of the presence of substructures}\label{sec:statAn_subs}
We now study  the effect of the presence of substructures on the potential and HE mass reconstructions. We investigate this effect on the entire cluster sample.
The results are shown in Fig.~\ref{fig:statAn_SubEffect} for the effects on the HE mass, the X-ray and the tSZ potentials. As expected from our investigations on two clusters in Sect.~\ref{sec:testcases} (see Sect.~\ref{sec:testcases_sub}), we observe that:
\begin{itemize}
\item{}The presence of substructures affects  the HE mass estimate only slightly (see Fig.~\ref{fig:statAn_MassSub}), with a difference in the bias of a few percent.
\item{}This effect is large on the X-ray potential (see Fig.~\ref{fig:statAn_PotXRSub}): the 31\% bias reduces to 15\% once the substructures are removed. A similar trend is observed for the spread of the distribution, with a reduction from 38\% to 17\%. This shows the importance of removing substructures prior to reconstructing X-ray potentials.
\item{}The effect due to the presence of substructures is small on the tSZ potential (as expected due to the linear dependence of the pressure on the gas density, see Fig.~\ref{fig:statAn_PotSZSub}).
\end{itemize}
We will further discuss these results in Sect.~\ref{sec:disc}.

 \begin{table*}
\begin{center}
\begin{tabular}{|l|c|c|c|l|}\hline
&M$_{HE}$& $\Phi_{tSZ}$ & $\Phi_{XR}$  &Section and Figures\\\hline
All clusters in & (0.13, 0.24) & (0.06, 0.15) & (0.15, 0.17)$_{\text{true, thermal}}$ &Sect.~\ref{sec:statAn_dynstate}\\
any dyn. state & & & &Fig.~\ref{fig:statAn_dynstateAll} and ~\ref{fig:statAn_embolo_allcl}\\\hline
HE effect& (0.03, 0.17)$_{\text{total}}$   & (-0.03, 0.13)$_{\text{total}}$ & (-0.06, 0.14)$_{\text{total, bolo}}$ &Sect.~\ref{sec:statAn_HE}\\
      & (0.14, 0.19)$_{\text{thermal}}$ & (0.08, 0.15)$_{\text{thermal}}$ & (0.02, 0.11)$_{\text{thermal, bolo}}$ & Fig.~\ref{fig:statAn_HEeffectHEMass},~\ref{fig:statAn_HEeffectPotSZ}, and~\ref{fig:statAn_embolo_relcl} \\\hline
Polytropic & & (0.06, 0.19)$_{\Delta\Gamma=0.03}$ & &Sect.~\ref{sec:statAn_Potpoly}\\
stratification & &(0.06,  0.23)$_{\Delta\Gamma=0.06}$ & & \\\hline                                
Presence of & &  & (-0.06, 0.14)$_{\text{bolo, total}}$ &Sect.~\ref{sec:statAn_xraybrem}\\
emission lines         & & & (0.02, 0.07)$_{\text{true, total}}$&Fig.~\ref{fig:statAn_embolo_relcl} \\\hline

Presence of & (0.15, 0.25)$_{\text{with sub}}$& (0.11, 0.19)$_{\text{with sub}}$&(0.31, 0.38)$_{\text{with sub}}$ & Sect.~\ref{sec:statAn_subs}\\
substructures           & (0.13, 0.24)$_{\text{no sub}}$ & (0.06, 0.15)$_{\text{no sub}}$& (0.15, 0.17)$_{\text{no sub}}$ &Fig.~\ref{fig:statAn_SubEffect} \\\hline

\end{tabular}
\caption{We report here the mean and root mean square (\textit{rms}) of the distributions of relative residuals as a measure of their \textit{bias} and \textit{scatter}.  These values have been obtained for the entire cluster sample, except for the effects due to the HE and to the presence of emission lines, for which we used the relaxed cluster sample. To single out how the X-ray potential is affected by the HE assumption  we use here  \textit{emissivity corrected} potentials (indicated with the subscript \textit{bolo}, to be distinguished from the \textit{true} X-ray emissivity). Likewise, to discriminate the effect due to the presence of emission lines from the HE effect,  the results reported here are derived for the total pressure (indicated with the subscript \textit{total}, to be distinguished from the \textit{thermal} pressure).  For the effect due to the polytropic stratification relation, we report the bias obtained with the fitted value of $\Gamma$ and the scatter obtained from the increased uncertainty on $\Gamma$:  $\Delta\Gamma=0.03$ and 0.06.}
\label{tab:syst}
\end{center}
\end{table*}

\section {Discussion}\label{sec:disc}

We will review in this section the systematics of the reconstructed potentials and HE mass, derived in Sect.~\ref{sec:statAn} and reported in Tab.~\ref{tab:syst}. As outlined in Sect.~\ref{sec:pot_method}, the method to reconstruct the X-ray and tSZ potentials relies on the assumption  that the clusters are in hydrostatic equilibrium (Eq.~[\ref{eq:eqHydro}]), that their pressure and gas density profiles follow a polytropic relation of constant polytropic index (Eq.~[\ref{eq:eqPoly}]), and that clusters are spherical systems. For the reconstruction from the X-ray emissivity, we also assume that the X-ray signal is dominated by the bolometric Bremsstrahlung emissivity. We will now discuss the effect of each of these assumptions on the potential reconstruction, and compare them, when appropriate,  with the systematics on the HE mass.
\begin{enumerate}
\item{}\textit{Effect of the cluster dynamical state} (Sect.~\ref{sec:statAn_dynstate}):
\\\\We investigated the reconstruction of the tSZ potential and HE mass on the entire cluster sample (85 clusters) and on the relaxed clusters (32 clusters). We observed that the tSZ potential is reconstructed with a smaller bias and scatter than the HE mass for clusters in any dynamical state (see Fig.~\ref{fig:statAn_dynstateAll}). This implies that the cluster potential is less affected by the overall reconstruction method than the HE mass. This is a promising result, which demonstrates that cluster potentials could indeed be a good alternative to cluster masses in cosmological studies. We will investigate the cosmological implications resulting from this low bias and variance in an upcoming paper.\\

\item{}\textit{Effect of the HE assumption} (Sect.~\ref{sec:statAn_HE}):
\\\\Using the relaxed cluster sample, we studied how the HE assumption affects the HE mass, as well as the X-ray, and tSZ potentials (see Fig.~\ref{fig:statAn_HEEffect}). 
As the deviations from the steady-state assumptions are expected to be small for these clusters, their HE masses and reconstructed potentials are not expected to be significantly affected by the HE assumption, once the contribution of non-thermal pressure support is accounted for in the HE equation (Eq.~[\ref{eq:eqHydro}]).  
\\\\We observed that, 
\begin{itemize}
\item Once the contribution from non-thermal pressure is taken into account,  the bias in the HE mass reduces from 14\% to 3\% (see Fig.~\ref{fig:statAn_HEeffectHEMass}).  It is interesting that a positive bias of a few percent remains even when we use the total pressure. This may be due to the assumption of spherical symmetry or alternatively, to the steady-state assumptions. We will elaborate further on this result in our upcoming paper, which is dedicated to the effects due to the cluster triaxiality.  \\

\item{} For the X-ray potentials, the bias reduces from 15\% to 2\%  once the contribution of non-thermal pressure support is accounted for in the reconstructions. These results have been obtained using the true X-ray emissivity of the clusters. For completeness, we also considered the HE effect on  the \textit{emissivity corrected} X-ray potentials (as defined in Sect.~\ref{sec:statAn_xraybrem}) and reported the bias and scatter of these distributions in Tab.~\ref{tab:syst}.  Interestingly, we found that the \textit{emissivity corrected} X-ray potentials reconstructed with the thermal pressure (not shown in a figure) has a low bias of 2\%, and that this bias becomes negative (by about $6\%$) once we take the contribution of the non-thermal support into account (see Fig.~\ref{fig:statAn_embolo_relcl}). This could be due to the substructure removal technique, which may be too strict and exclude pixels which belong to the cluster (as discussed in the point 6 below). Nevertheless, this relaxed cluster sample does not seem to be significantly affected by the HE assumption once we corrected for the assumption on the X-ray emissivity.  \\
\item{} With a bias of 8\%, the tSZ potentials reconstructed for the thermal pressure perform already quite well (see Fig.~\ref{fig:statAn_HEeffectPotSZ}). Interestingly,  the potentials reconstructed using the total pressure overestimates the true potential by 3\% . This could be due to the substructure removal technique (as we also noticed for the \textit{emissivity corrected} X-ray potentials). \\
\end{itemize}



\item{}\textit{Effect of the polytropic stratification assumption} (Sect.~\ref{sec:statAn_Potpoly}):
\\\\Cluster potentials are reconstructed in this work assuming a polytropic stratification. This assumption is only required to express the potential in a simple form  (see Eq.~[\ref{eq:recX}] and ~[\ref{eq:recSZ}]). We found that the potentials reconstructed for uncertainties on $\Gamma$  of a few percent ($\Delta\Gamma_{average}=0.03$) have a smaller scatter (19\%) than the HE mass (24\%). In the study presented here we fitted the value of $\Gamma$ on an optimized radial range (within 0.2$\text{R}_{500}$ and $\text{R}_{500}$). Such an optimization is possible on real data and is expected to yield results similar to the ones presented here \citep[see e.g.][1.21+/- 0.03]{eckert13}.

For completeness, we also compare our potentials to potentials reconstructed without assuming a polytropic stratification relation \citep[HE potential, as in][]{lau11}. This HE potential only relies on the HE and spherical symmetry assumptions, and is reconstructed directly from pressure and density profiles. We show in Fig.~\ref{fig:statAn_CompaPotLau11} the relative residuals for the HE potentials, the tSZ and X-ray potentials, and the HE mass. As we can see, all these potentials have been reconstructed with a bias smaller than that of the HE mass. In addition, if we are able to set tight constraints on the value of $\Gamma$, our method has the ability to yield more accurate cluster potentials reconstructions than a method that does not assume a polytropic stratification relation. 
\\

\item{}\textit{Effect of the contribution of emission lines to the X-ray emissivity} (Sect.~\ref{sec:statAn_xraybrem}):
\\\\We examined the effect of the assumed bolometric Bremsstrahlung emissivity on the X-ray potential. This assumption is required to express the X-ray potential in the simple form given in Eq.~[\ref{eq:recX}].  Our results show that this effect increases the bias from 9\% to 15\% for the clusters in any dynamical state. Considering the relaxed cluster sample, we also found that the reconstruction using the thermal pressure returns a bias of only 2\%. It implies that the assumed bolometric Bremsstrahlung emissivity does not affect significantly the X-ray reconstructed potentials of this relaxed cluster sample (see Fig.~\ref{fig:statAn_embolo_relcl}). \\

\item{}\textit{Effect of the presence of substructures} (Sect.~\ref{sec:statAn_subs}):
\\\\We investigated how the presence of substructures affects the HE mass and the reconstructed potentials. We observed that the effect is of a few percent for the HE mass and the tSZ potential, but that it becomes larger for the X-ray potential (see Fig.~\ref{fig:statAn_SubEffect}). While we expect the tSZ potential to be less affected than the X-ray potential, due to their different dependency on the gas density, the effect on the HE mass is interesting and similar to the result we obtained from the analysis of the two clusters CL77 and CL135 (in Sect.~\ref{sec:testcases}). \\
\item{}\textit{Notes on the effect due to the spherical symmetry assumption}:
\\Apart from the uncertainties in neglecting triaxiality, we also found that, for non-spherical clusters, the procedures used to remove substructures are expected to introduce additional uncertainties in the reconstructions (see the introduction of Sect.~\ref{sec:syst} and Fig.~\ref{fig:Xsigma}). We summarize below different examples we encountered and that may be caused, at least partially, by the triaxiality of the simulated clusters. The detailed investigation of this effect  will be reported in our upcoming study (Tchernin et al., in prep). \\
\begin{itemize}
\item{} We reconstructed  the HE mass and the potentials of the relaxed clusters sample using the total pressure, and found out that a non-zeros bias of  few percent is still remaining (see Fig.~\ref{fig:statAn_HEEffect} and Fig.~\ref{fig:statAn_embolo_relcl});\\
\item{} Studying the effect of the HE assumption, we observed a difference in bias and scatter between the reconstructed tSZ and X-ray potentials (Fig.~\ref{fig:statAn_HEEffect}). This could be due to the assumed bolometric emissivity (as seen in Fig.~\ref{fig:statAn_embolo_relcl}), combined to the different sensitivity of these two probes to the substructure removal technique. However, it could also highlight the different origin of these two signals. We will be able to further investigate  the source of this discrepancy once we take the triaxiality of the clusters into account in our potential reconstructions (as reported in Tchernin et al., in prep). \\
\item{}  For the relaxed cluster CL135, we observed that the median tSZ potential is performing worse than the mean tSZ potential (see Fig.~\ref{fig:pot}).  Typically, pressure profiles are expected to be only slightly affected by the presence of substructures due to their linear dependence in the density \citep[see e.g.,][]{tchernin16}. However, we observed that for this cluster, the structure removed (median) and original (mean) pressure profiles differ (see  Fig.~\ref{fig:pressure}). The relative residuals, shown in the bottom panel of this figure, even point out that the discrepancy between these two profiles monotonically increases with the concentric radius. This  hints toward an effect of the assumption of spherical symmetry. Indeed, while substructures have a local effect on profiles, the triaxiality of clusters is expected to affect the entire cluster range. This result  is also consistent with our findings shown in Fig.~\ref{fig:triaxial_median}: the median profiles extracted in spherical shells are biased low for clusters with triaxial shapes. \\
\item{}  In the analysis presented in Sect.~\ref{sec:statAn}, we used the 3.5$\sigma$ criterion. In Fig.~\ref{fig:statAn_SubMedian} we show the reconstruction of the X-ray potential and HE mass derived for the two techniques: the median (\textit{Method 1}) and the 3.5-$\sigma$ criterion (\textit{Method 2}).
Interestingly, these two approaches  affect the HE mass and X-ray potential in opposite ways. This result can be due to their different sensitivity to the presence of substructures (see Sect.~\ref{sec:statAn_subs}) combined to the spherical symmetry  assumption.
We also note that the scatter in the median X-ray potential is noticeable and raises the question if the median technique is adequate to reconstruct the X-ray potential of triaxial clusters. \\
\end{itemize}

\end{enumerate}


\section {Conclusion}\label{sec:conc}
Assuming spherical symmetry, we studied the systematics  of the X-ray and tSZ reconstructed potential, and compared them to that of the HE mass estimates. We based our analysis on the 85 clusters of the Omega500 simulations, which are in various dynamical state.
Our results show that the recovery of the potential performs better than that of the HE mass.
\\\\In particular, our results show that 
\begin{itemize}
    \item The bias and scatter in the distribution of HE mass (13\%, 24\%) are larger than those in the distribution of tSZ potential (6\%, 15\%) for clusters in different dynamical states;\\
    \item Considering the relaxed clusters sample, the hydrostatic equilibrium assumption increases the bias in the HE mass from 3\% to 14\%, the bias in the tSZ potential from -3\% to 8\%, and the bias in the X-ray potential from 2\% to 15\%. However, it does not seem to affect significantly the spread of these distributions. The different level of systematics in the tSZ and X-ray potential reconstructions  could suggest that these two probes are sensitive to ICM components that are from different origin or in different physical states;\\
    \item Uncertainties on the polytropic index value comparable to the ones obtained in the real observations ($\Delta\Gamma_{average}=0.03$) generate a spread in the reconstructed tSZ potentials of 19\%. This is still smaller than the spread in the HE mass (24\%);\\
    \item The assumed bolometric Bremsstrahlung emissivity affects the bias of the distribution of X-ray reconstructed potentials from 9\% to 15\% (considering clusters in various dynamical state), but leaves its spread almost unchanged;\\
    \item{}The presence of substructures affects the HE mass and the reconstructed potentials differently. We observed that removing these density inhomogeneities is crucial for the reconstruction of X-ray potentials (the bias  and scatter (31\%,38\%) reduce to (15\%, 17\%) once the substructures are removed). This effect is, however, less significant for the HE mass and the tSZ potential.
    \end{itemize}
    
     \noindent Assuming spherical symmetry, we investigated two different substructure removal techniques: the \textit{median} and the X-$\sigma$ approaches (as introduced in Sect.~\ref{sec:syst}). We observed that by applying a too strict density threshold for the X-$\sigma$ criterion, we exclude fluctuations of matter which actually belong to the cluster (as in Fig.~\ref{fig:Xsigma_map}). We also noticed that taking the median value of the density distribution of a triaxial cluster, biases low the resulting profiles (Fig.~\ref{fig:triaxial_median}). These results hint at the effects of the spherical symmetry assumption. We will investigate them in more detail in our upcoming paper (Tchernin et al., in prep).
\\\\Based on the low variance and bias we were able to achieve with our  reconstructed potentials and on the large number of expected advantages of the gravitational potential with respect to the cluster mass (see Sect.~\ref{sec:adv}), we argue here that the cluster potential offers a promising way to characterize galaxy clusters in cosmological studies. 
We are currently investigating the final impact on cosmological studies. We will report our results and the  scatter in the underlying \textit{observable-mass} and \textit{observable-potential} scaling relations in an upcoming, dedicated study.

\acknowledgements{CT acknowledges the financial support from the Deutsche Forschungsgemeinschaft under BA 1369 / 28-1 and from the Swiss National Science Foundation under P2GEP2\_159139. We thank Daisuke Nagai for kindly giving us the permission to use the data from the {\em Omega500} simulation. The {\em Omega500} simulations were performed and analyzed at the HPC cluster {\em Omega} at Yale University, supported by the facilities and staff of the Yale Center for Research Computing. }

\newpage
\appendix
\section{Additional figures}

\begin{figure}[h!]
  \centering
  \includegraphics[width=0.7\columnwidth]{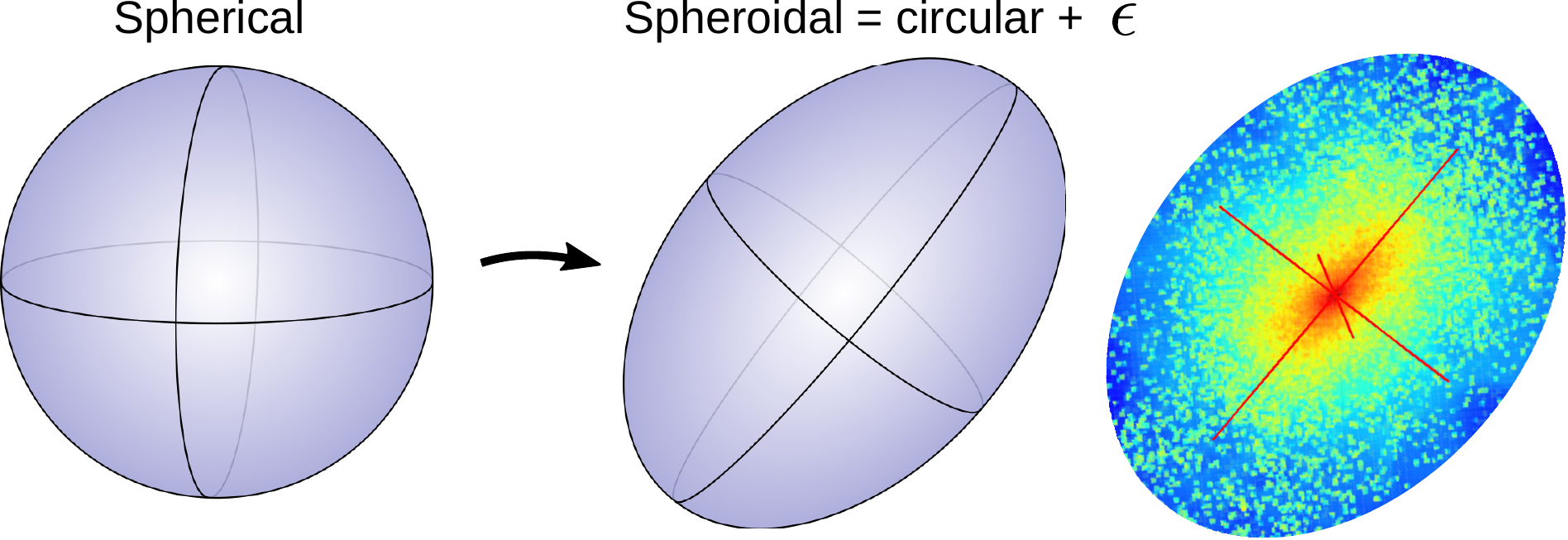}
\caption{Sketch of the spherical (left panel) and the spheroidal (for a 3D body with: c=b$<$a, middle panel) cluster morphologies. The right panel shows the gas and DM distribution of the cluster relaxed CL135 (see Sect.~\ref{sec:testcases}) overlaid to the spheroidal case (the axes show the eigenvectors of the gas distributions).}
\label{fig:sketch}
\end{figure}

\begin{figure}[h!]
\begin{subfigure}{.5\textwidth}
  \centering
  \includegraphics[width=\columnwidth,height=0.7\columnwidth,angle=0]{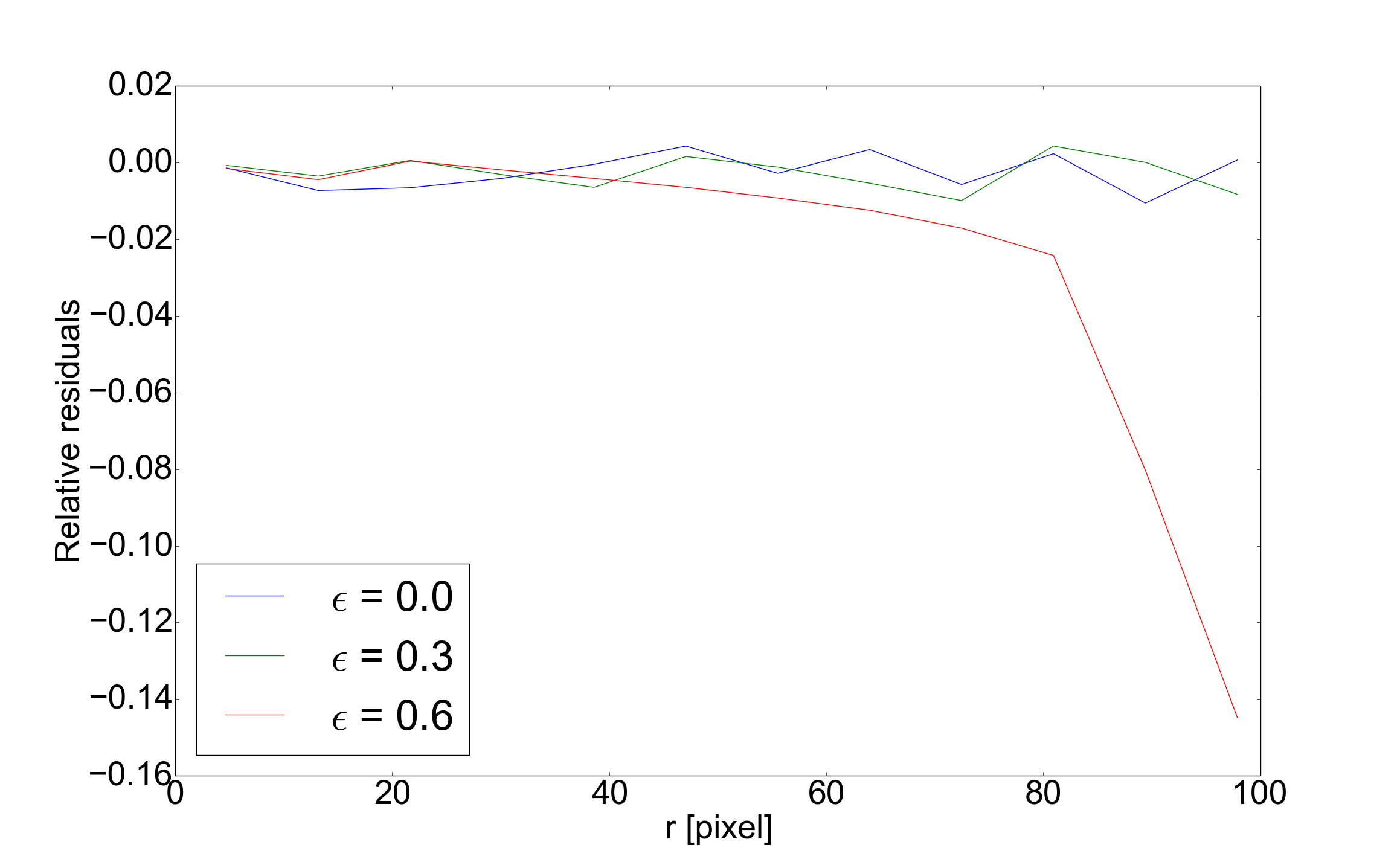}
  \caption{}
   \label{fig:triaxial_median}
\end{subfigure}
\begin{subfigure}{0.5\textwidth}
  \centering
  \includegraphics[height=0.7\columnwidth,angle=0]{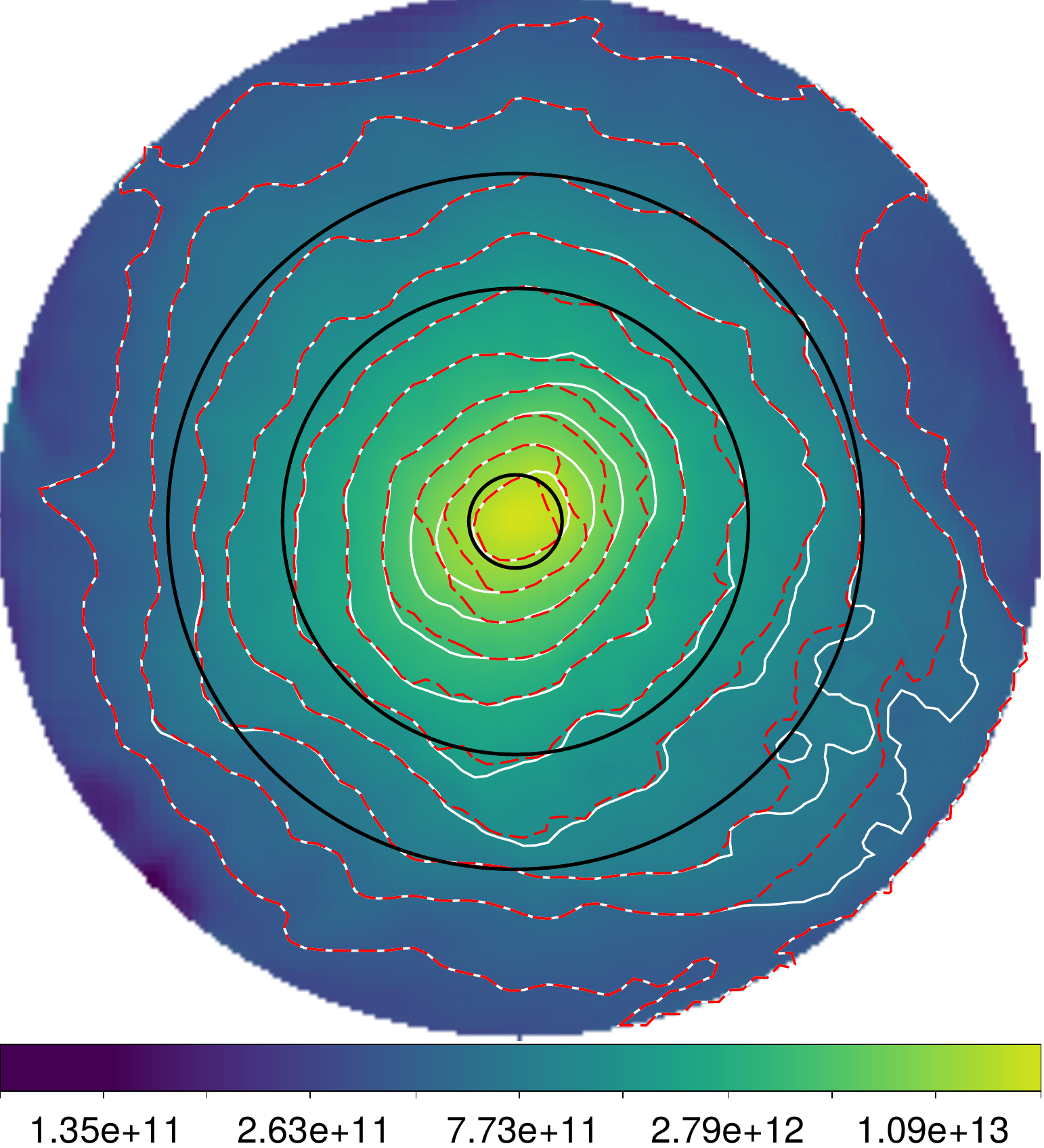}
  \caption{}
   \label{fig:Xsigma_map}
\end{subfigure}
\begin{subfigure}{.5\textwidth}
  \centering
  \includegraphics[width=\columnwidth,height=0.7\columnwidth,angle=0]{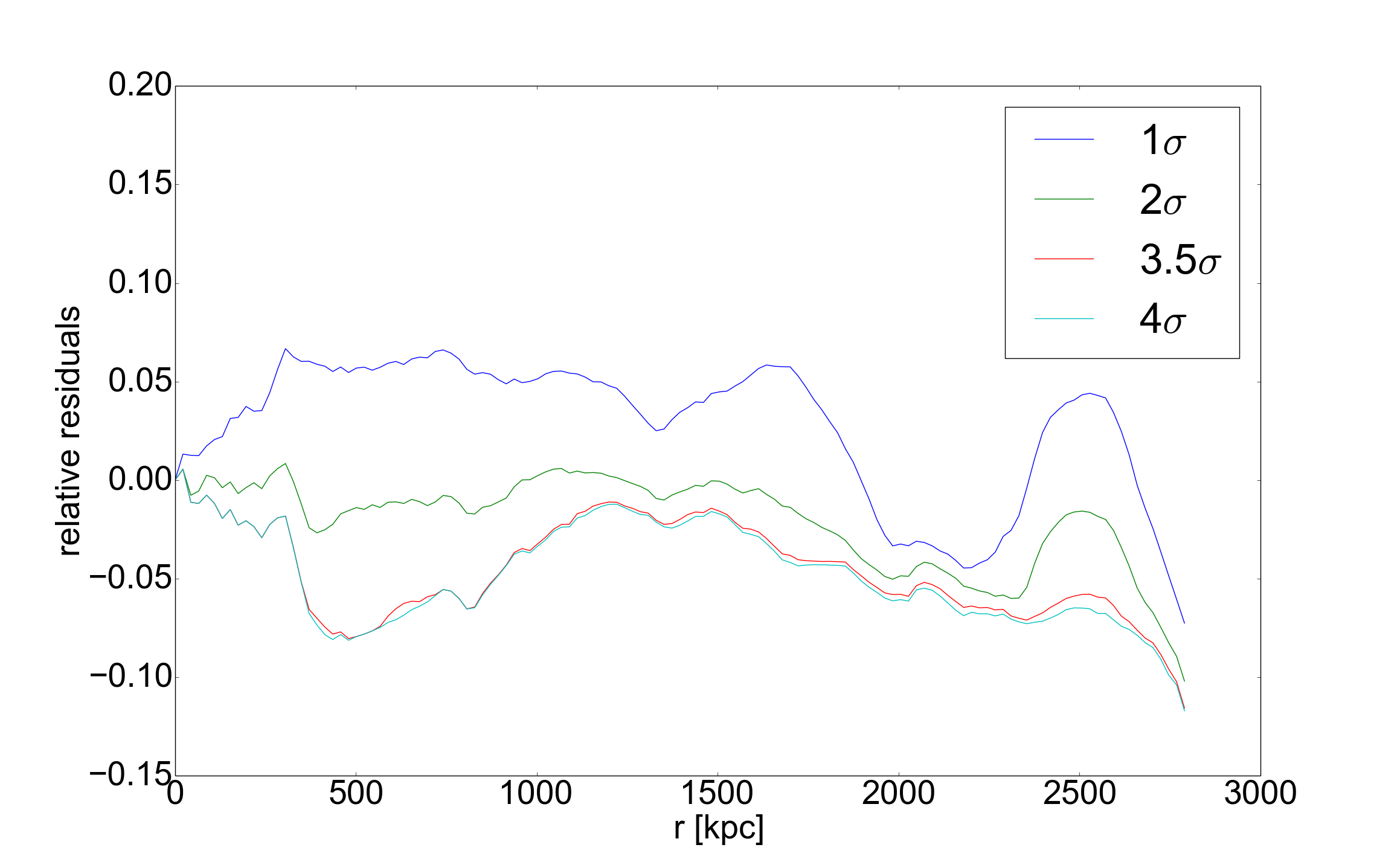}
  \caption{}
   \label{fig:Xsigma_profile}
\end{subfigure}
\caption{Illustration of the two substructure removal methods introduced in Sect.~\ref{sec:syst}:
a) \textit{Method 1}: Median technique applied to spheroidal clusters of increasing ellipticity ($\epsilon=0,0.3$ and 0.6). With no substructures and assuming spherical symmetry. The relative residuals are computed as: (median-mean)/median.
b) \textit{Method 2}: Effect of the threshold value X$\sigma$ on the projected gas density of the cluster CL135: the isocontours  are shown in red for X=3.5 and in white for X=2. The black circles indicate $0.2\text{R}_{500}$, $\text{R}_{500}$ and $\text{R}_{200}$. The units are expressed in [M$_{\odot}$/Mpc$^3$]; c) relative residuals between the profiles obtained with these two methods, computed as (\textit{Method 1}-\textit{Method 2})/\textit{Method 1}, for different threshold X$\sigma$ values (X= 1, 2, 3.5 and 4).}
\label{fig:Xsigma}
\end{figure}

\begin{figure}[h!]
\begin{subfigure}{.5\textwidth}
  \centering
  \includegraphics[width=\columnwidth,height=0.7\columnwidth,angle=0]{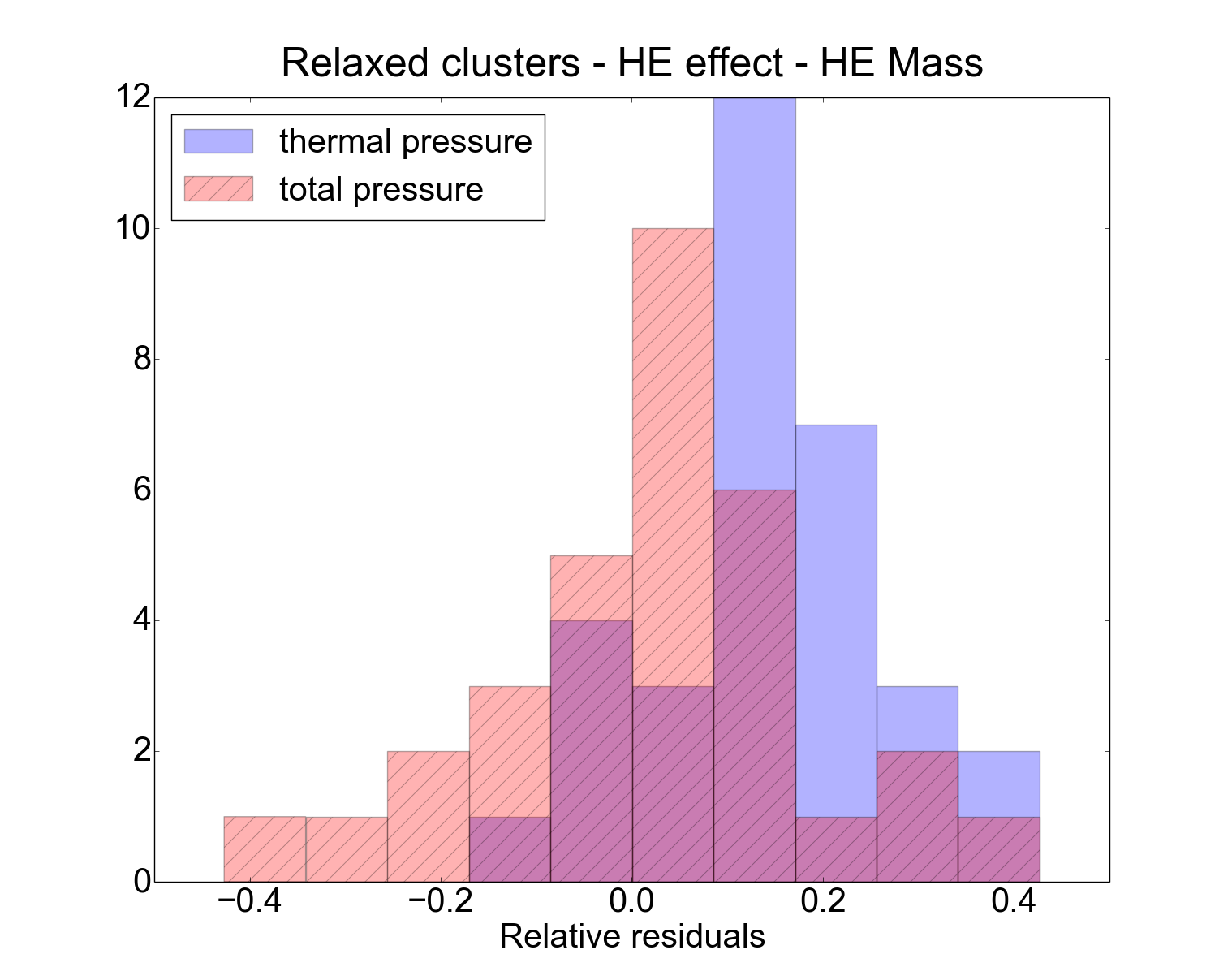}
  \caption{}
   \label{fig:statAn_HEeffectHEMass}
\end{subfigure}
\begin{subfigure}{.5\textwidth}
  \centering
  \includegraphics[width=\columnwidth,height=0.7\columnwidth,angle=0]{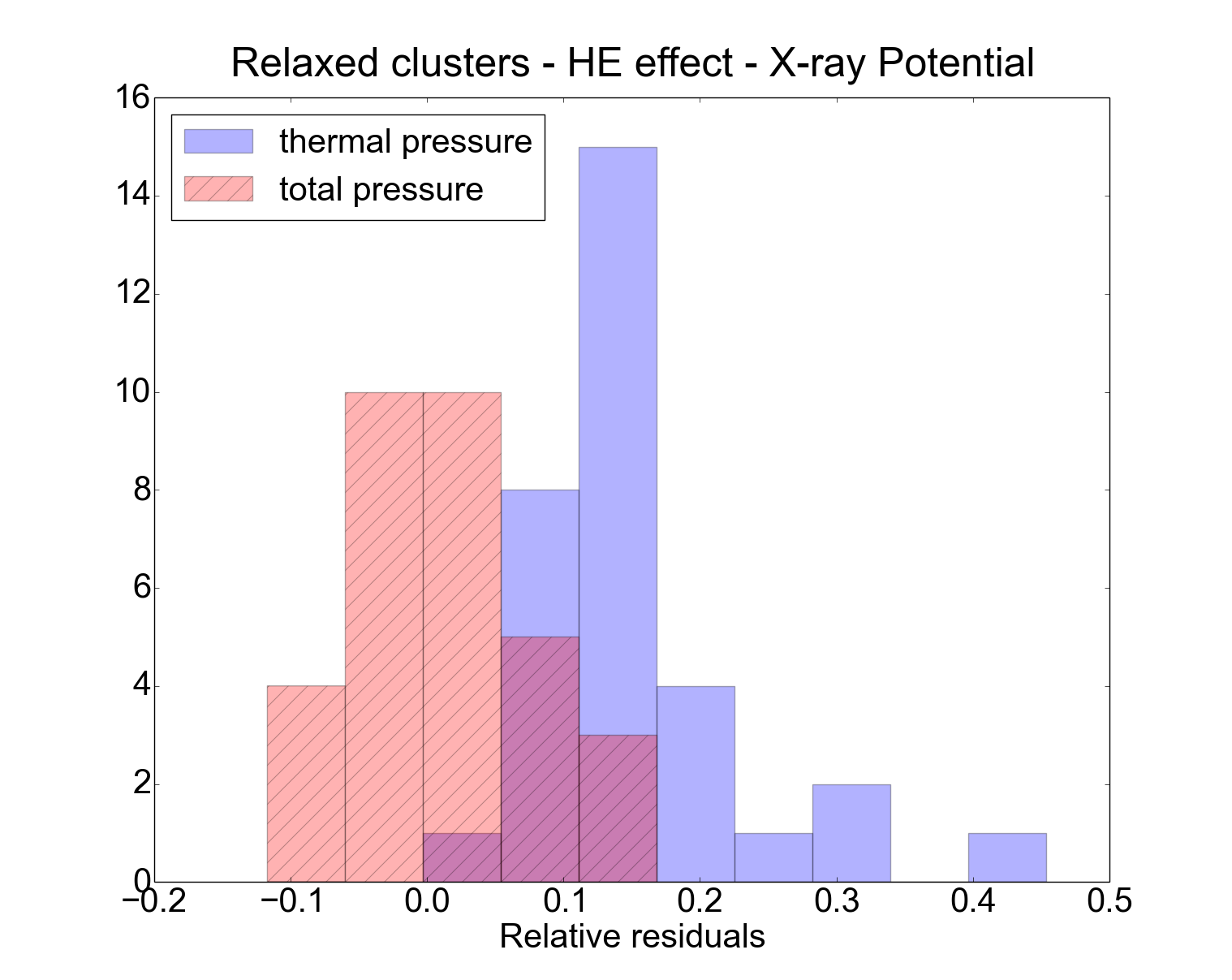}
  \caption{ }
   \label{fig:statAn_HEeffectPotXR}
\end{subfigure}
\begin{subfigure}{.5\textwidth}
  \centering
  \includegraphics[width=\columnwidth,height=0.7\columnwidth,angle=0]{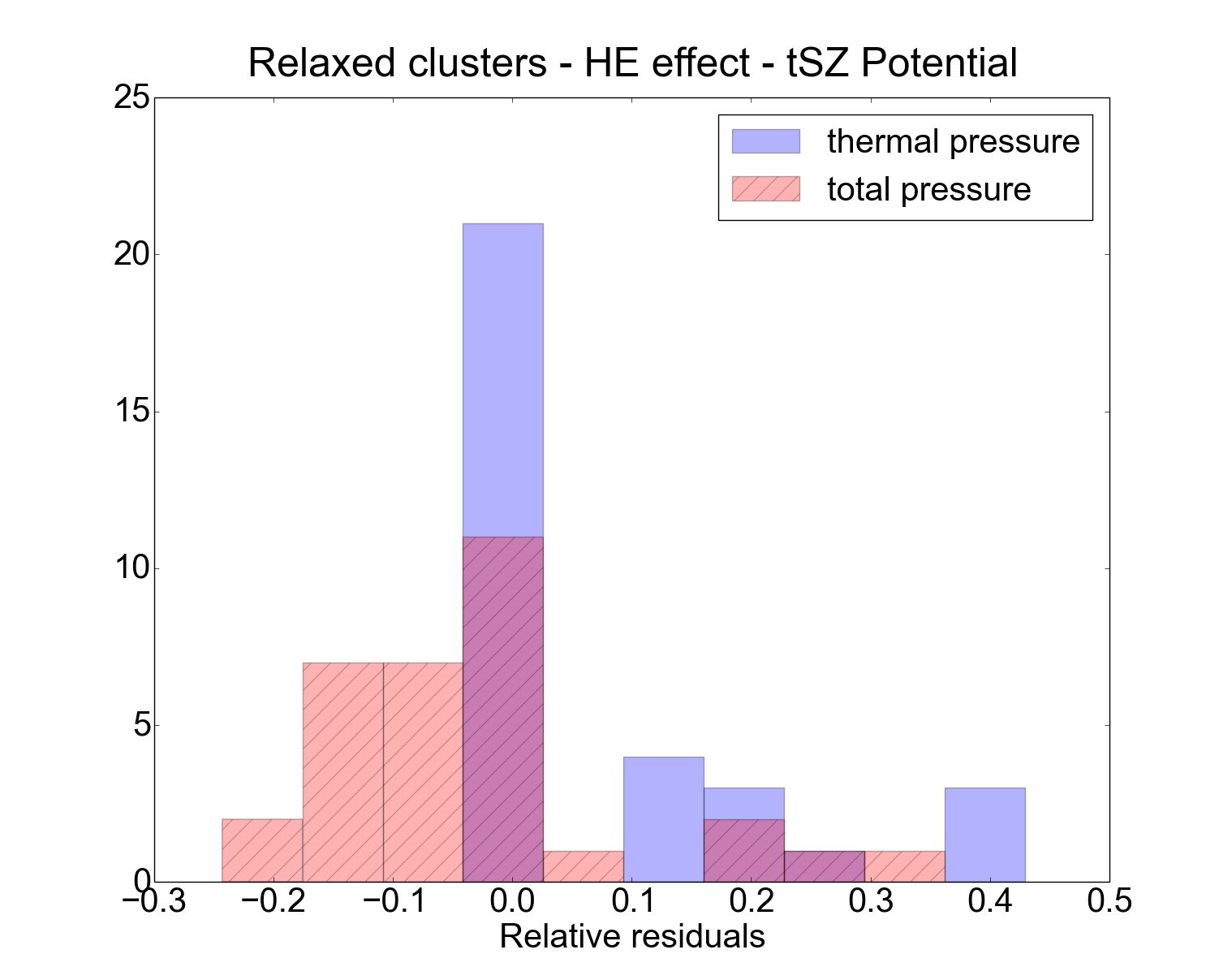}
  \caption{}
   \label{fig:statAn_HEeffectPotSZ}
\end{subfigure}
\caption{Relative residuals - effect of the HE assumption, expressed as (\textit{bias}, \textit{scatter}) and evaluated at $\text{R}_{500}$ for: a) the HE mass, $M_{total}$=(0.03, 0.17); $M_{thermal}$=(0.14, 0.19): b) the X-ray potential, $\Phi_{total}$=(0.02, 0.07); $\Phi_{thermal}$=(0.15, 0.17); c) the tSZ potential, $\Phi_{total}$=(-0.03, 0.13); $\Phi_{thermal}$=(0.08, 0.15).}
\label{fig:statAn_HEEffect}
\end{figure}


 \begin{figure}[h!]
\begin{subfigure}{.5\textwidth}
  \centering
  \includegraphics[width=\columnwidth,height=0.7\columnwidth,angle=0]{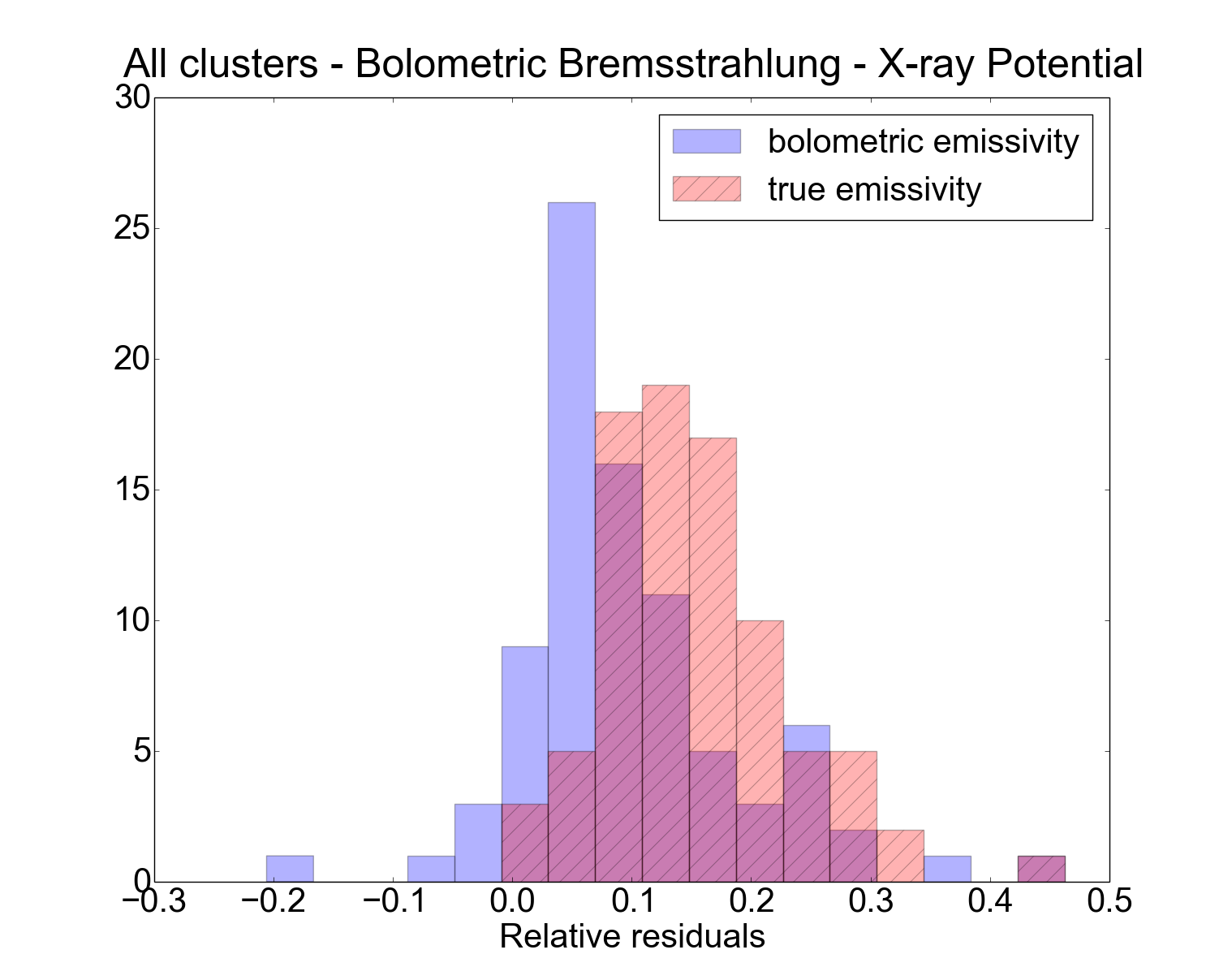}
  \caption{}
   \label{fig:statAn_embolo_allcl}
\end{subfigure}
\begin{subfigure}{.5\textwidth}
  \centering
  \includegraphics[width=\columnwidth,height=0.7\columnwidth,angle=0]{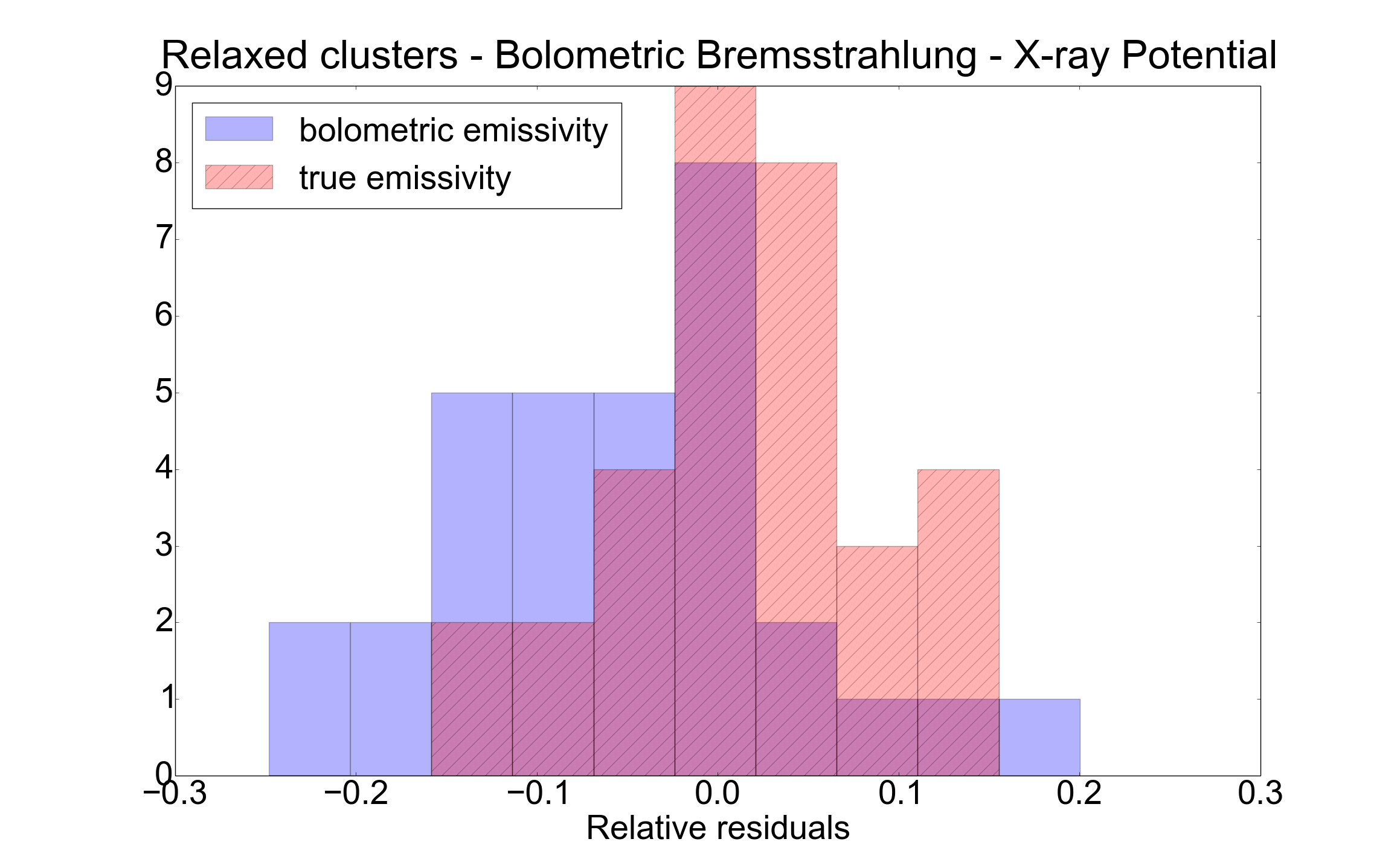}
  \caption{ }
   \label{fig:statAn_embolo_relcl}
\end{subfigure}
\caption{Relative residuals - effect of the assumed X-ray bolometric Bremsstrahlung radiation, expressed as (\textit{bias}, \textit{scatter}) and evaluated at $\text{R}_{500}$ for: a) The entire cluster sample, using the \textbf{thermal pressure}:  $\Phi_{true}$=(0.15, 0.17), $\Phi_{bolo}=(0.09, 0.14)$; b) For the relaxed clusters sample, using the \textbf{total pressure}: $\Phi_{true}$=(0.02, 0.07), $\Phi_{bolo}=(-0.06, 0.14)$.}
\label{fig:statAn_embolo}
\end{figure}

\begin{figure}[h!]
\begin{subfigure}{.5\textwidth}
  \centering
  \includegraphics[width=\columnwidth,height=0.7\columnwidth,angle=0]{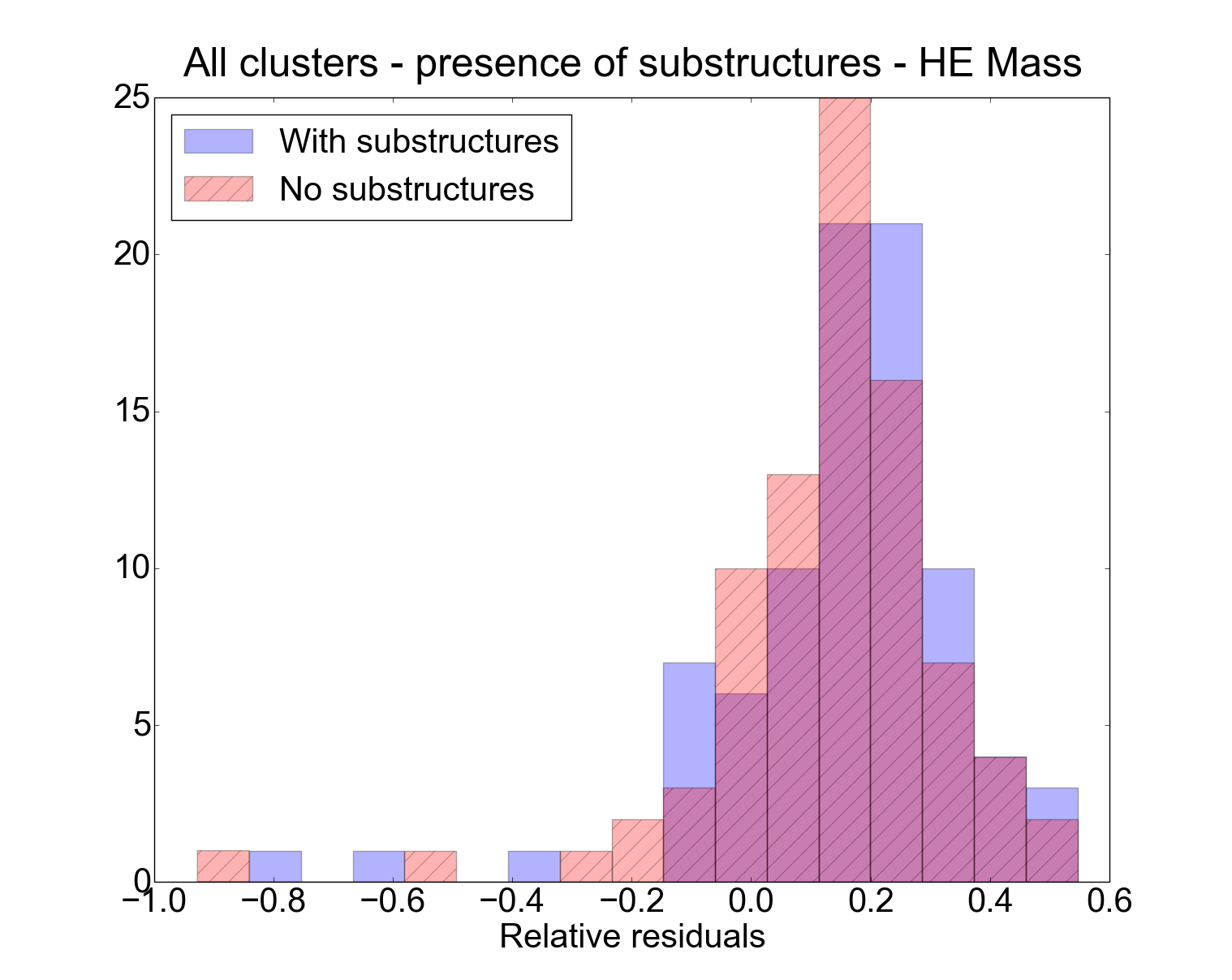}
  \caption{}
   \label{fig:statAn_MassSub}
\end{subfigure}
\begin{subfigure}{.5\textwidth}
  \centering
  \includegraphics[width=\columnwidth,height=0.7\columnwidth,angle=0]{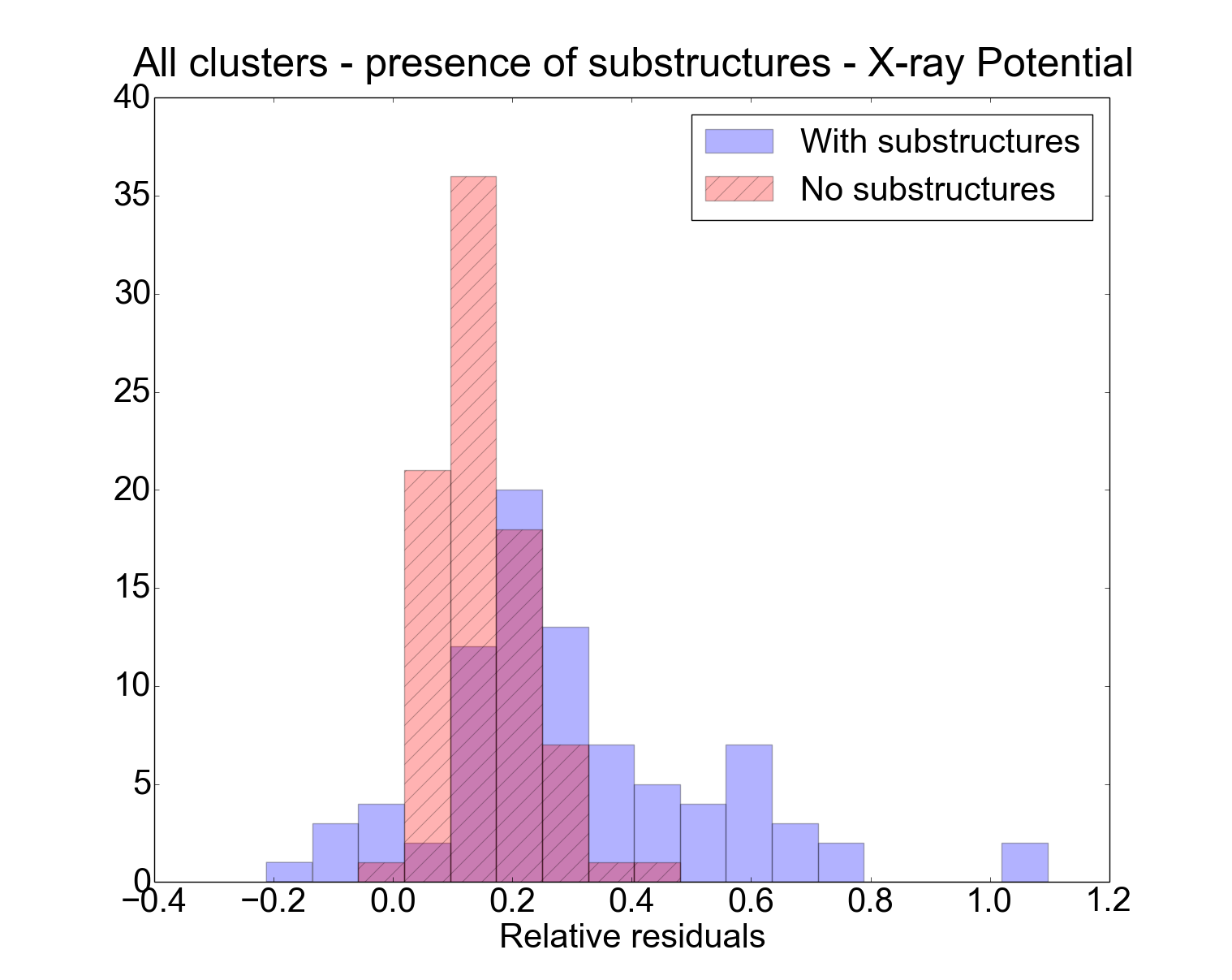}
  \caption{ }
   \label{fig:statAn_PotXRSub}
\end{subfigure}
\begin{subfigure}{.5\textwidth}
  \centering
  \includegraphics[width=\columnwidth,height=0.7\columnwidth,angle=0]{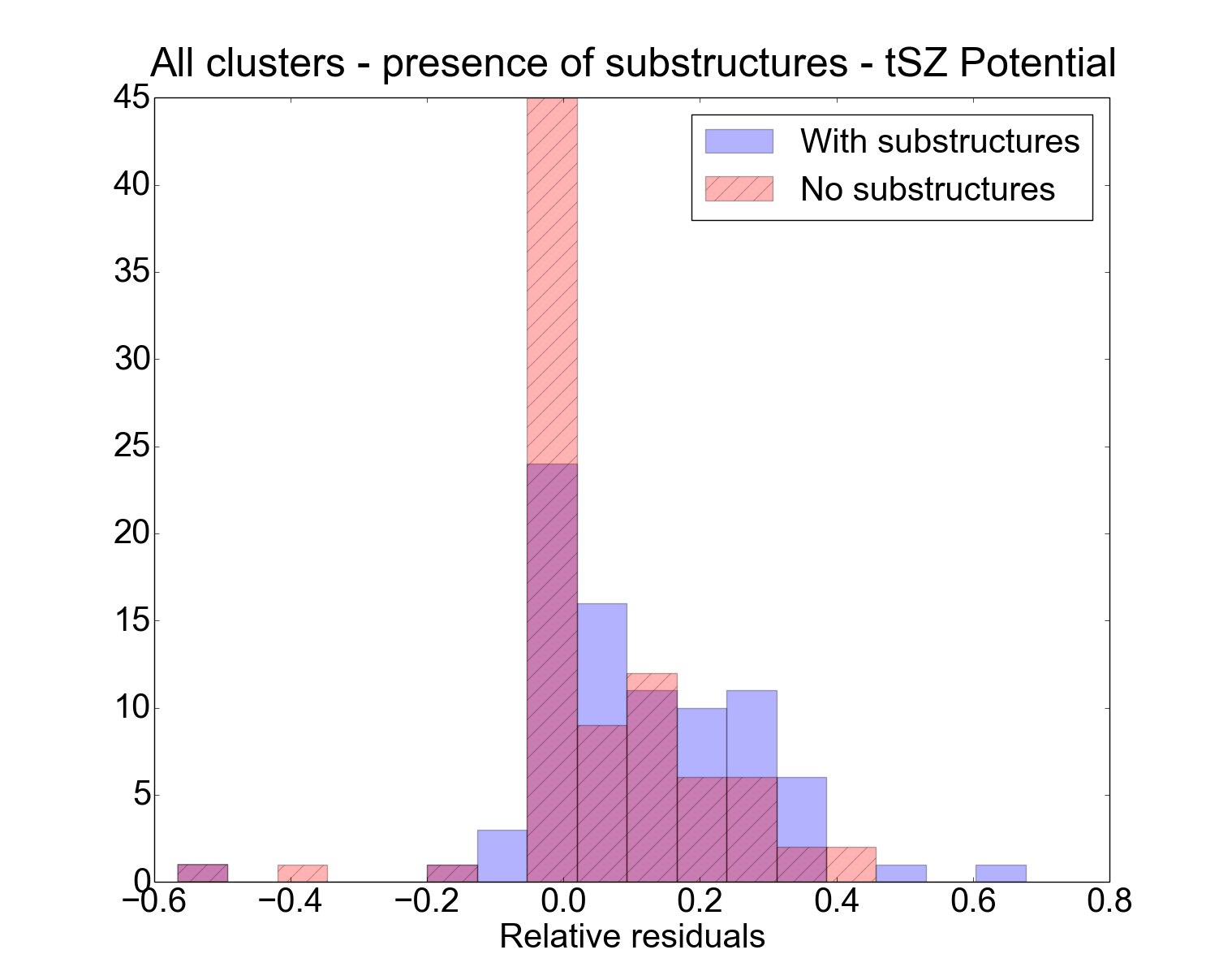}
  \caption{}
   \label{fig:statAn_PotSZSub}
\end{subfigure}
\caption{Relative residuals - effect of substructures, expressed as (\textit{bias}, \textit{scatter}) and evaluated at $\text{R}_{500}$ for: a) the HE mass, $M_{sub}$=(0.15,0.25); $M_{nosub}$=(0.13,0.24); b) the X-ray potential, $\Phi_{sub}$=(0.31,0.38); $\Phi_{nosub}$=(0.15,0.17); c) the tSZ potential, $\Phi_{sub}$=(0.11,0.19); $\Phi_{nosub}$=(0.06,0.15).}
\label{fig:statAn_SubEffect}
\end{figure}

\begin{figure}
\begin{center}
  \includegraphics[width=\columnwidth,height=0.7\columnwidth,angle=0]{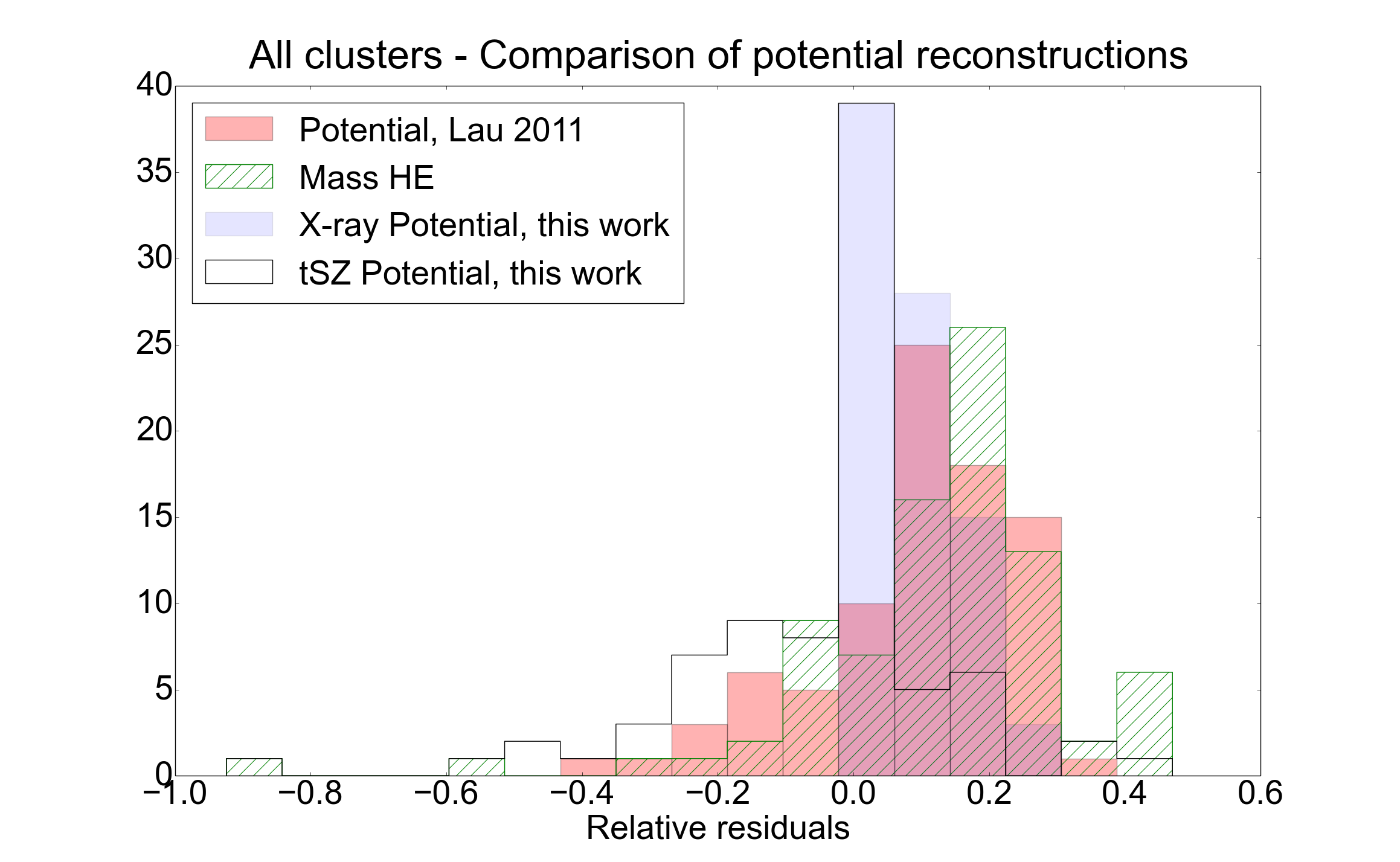}
  \caption{Relative residuals - Comparison of different methods to reconstruct cluster potentials. With the polytropic stratification assumption (this work): from X-rays (in blue), from the tSZ effect (in black contours), and without the polytropic stratification assumption \citep[in red,][]{lau11}. We show the results for the HE mass  for comparison (in hatched green).  The X-ray and tSZ potentials have been reconstructed for uncertainties on $\Gamma$ of a few per mill.}
   \label{fig:statAn_CompaPotLau11}
\end{center}
\end{figure}

\begin{figure}[h!]
\begin{subfigure}{.5\textwidth}
  \centering
  \includegraphics[width=\columnwidth,height=0.7\columnwidth,angle=0]{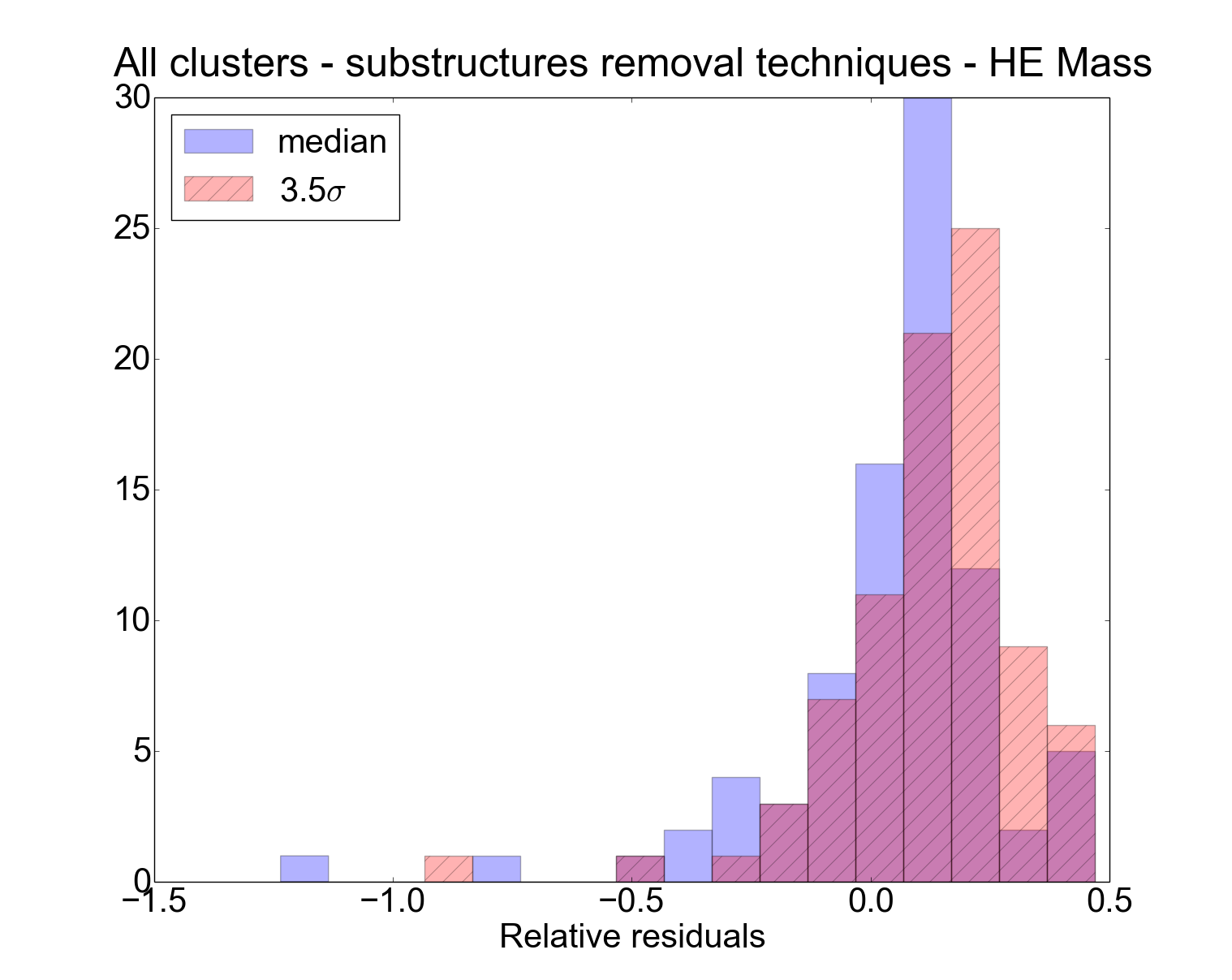}
  \caption{}
   \label{fig:statAn_SubEffectHEmedian}
\end{subfigure}
\begin{subfigure}{.5\textwidth}
  \centering
  \includegraphics[width=\columnwidth,height=0.7\columnwidth,angle=0]{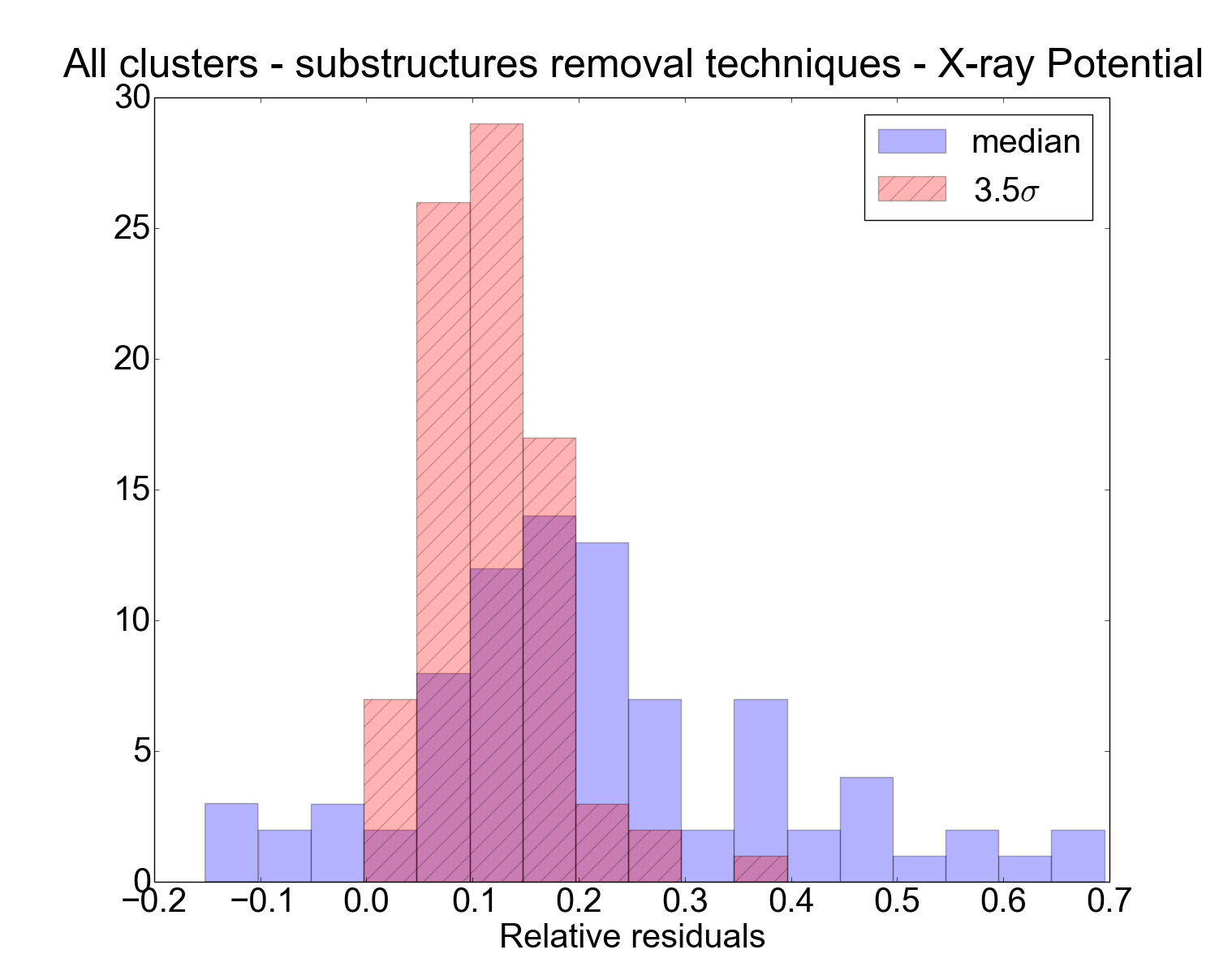}
  \caption{ }
   \label{fig:statAn_SubEffectXRmedian}
   \end{subfigure}
\caption{Relative residuals - Comparison of the substructure removal technique - 3.5$\sigma$ and median (as introduced in the first paragraph of Sect.~\ref{sec:statAn}), expressed as (\textit{bias}, \textit{scatter}) and evaluated at $\text{R}_{500}$ for: a) the HE mass: $M_{3.5\sigma}$=(0.13, 0.24), $M_{median}=(0.09, 0.25)$; b) the X-ray emissivity: $\Phi_{3.5\sigma}$=(0.15, 0.17), $\Phi_{median}=(0.21, 0.28)$.}
\label{fig:statAn_SubMedian}
\end{figure}

 \begin{figure}
  \centering
  \includegraphics[width=\columnwidth]{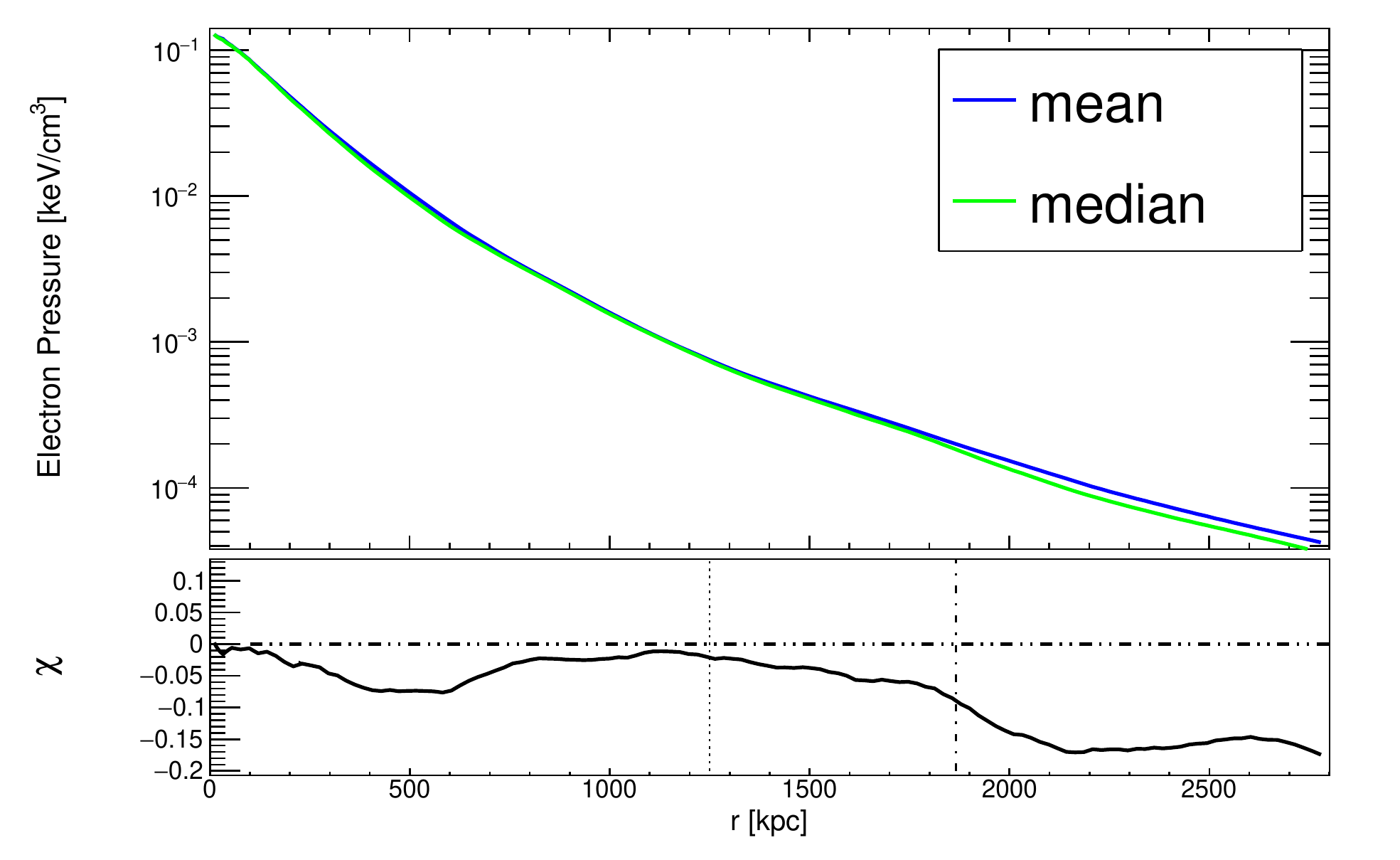}
\caption{Pressure profile of CL135. Top panel: mean (blue) and median (green) profiles derived in spherical shells. Bottom panel: relative residuals computed as ($\text{P}_{\text{median}}-\text{P}_{\text{mean}})/\text{P}_{\text{median}}$, with $\text{P}_{*}$ the corresponding profile value. The vertical lines represent $\text{R}_{500}$ and $\text{R}_{200}$.}
\label{fig:pressure}
\end{figure}

\end{document}